\documentclass[letter,11pt]{article}

\pdfoutput=1 % if your are submitting a pdflatex (i.e. if you have
\usepackage{jheppub} % for details on the use of the package, please
\usepackage{amsmath}
\usepackage{graphicx}
\usepackage{dcolumn}
\usepackage{bm}
\usepackage{bbm}
\usepackage{epsfig}
\usepackage{slashed}
\usepackage{amssymb}
\usepackage{color}
\usepackage[font=small]{caption}
\usepackage[font=small]{subcaption}
\usepackage{tabulary}

\def\lsim{\mathrel{\rlap{\lower4pt\hbox{\hskip1pt$\sim$}}
    \raise1pt\hbox{$<$}}}         %less than or approx. symbol
\def\gsim{\mathrel{\rlap{\lower4pt\hbox{\hskip1pt$\sim$}}
    \raise1pt\hbox{$>$}}}         %greater than or approx. symbol
\newcommand{\blue}{\textcolor{black}}
\newcommand{\red}{\textcolor{black}}

\newcommand{\be}{\begin{equation}}
\newcommand{\ee}{\end{equation}}
\newcommand{\bea}{\begin{eqnarray}}
\newcommand{\eea}{\end{eqnarray}}

\begin{document}

\vspace*{-2cm}
\title{\bf  Diboson Excesses Demystified in Effective Field Theory Approach}

 %%%%%%%%%%%%%%%%%%%%%%%%%%%%%%%%%%%%%%%%%%%%%%%%%%%%%%%%%%%%%%%%%
\author[\dagger]{Doojin Kim}
\affiliation[\dagger]{Department of Physics, University of Florida, Gainesville, FL 32611, USA}
  
\author[\ddagger]{, Kyoungchul Kong}
\affiliation[\ddagger]{Department of Physics and Astronomy, University of Kansas, Lawrence, KS 66045, USA}

\author[\sharp]{, Hyun Min Lee,}
\affiliation[\sharp]{Department of Physics, Chung-Ang University, Seoul 156-756, Korea}

\author[\ast\star]{, Seong Chan Park}
\affiliation[\ast]{Department of Physics and IPAP, Yonsei University, Seoul 120-749, Korea}
\affiliation[\star]{Korea Institute for Advanced Study, Seoul 130-722, Korea}

% e-mail addresses: one for each author, in the same order as the authors
\emailAdd{immworry@ufl.edu}
\emailAdd{kckong@ku.edu}
\emailAdd{hminlee@cau.ac.kr}
\emailAdd{sc.park@yonsei.ac.kr}

 %%%%%%%%%%%%%%%%%%%%%%%%%%%%%%%%%%%%%%%%%%%%%%%%%%%%%%%%%%%%%%%%%

\vspace*{2.0cm}
%\begin{abstract}
\abstract{
We study the collider implication of a neutral resonance which decays to several diboson final states such as $W^+W^-$, $ZZ$, and $Z\gamma$  via a minimal set of effective operators. We consider both CP-even and CP-odd bosonic states with spin 0, 1, or 2.
The production cross sections for the bosonic resonance states are obtained with the effective operators involving gluons (and quarks), and the branching fractions are obtained with the operators responsible for the interactions with electroweak gauge bosons. 
We demonstrate that each scenario allows for a broad parameter space which could accommodate the recently-reported intriguing excesses in the ATLAS diboson final states, and discuss how the CP states and spin information of the resonance can be extracted at the LHC run II. 
}
%\end{abstract}
\preprint{CETUP2015-013}

\keywords{LHC, diboson, resonance, effective operators}

\maketitle

%%%%%%%%%%%%%%%%%%%%%%%%%%%%%%%%%%%%%%%%%%%%%%%%%%%%%%%%%%%%%%%%%%%
\section{Introduction}  \label{sec:introduction}
%%%%%%%%%%%%%%%%%%%%%%%%%%%%%%%%%%%%%%%%%%%%%%%%%%%%%%%%%%%%%%%%%%%

Recently, the ATLAS collaboration has reported some excesses in searches for diboson resonances 
in the highly boosted final states with $W^+W^-$, $W^\pm Z$ and $ZZ$ at the 8 TeV LHC with 20.3 fb$^{-1}$ \cite{Aad:2015owa}. They have adapted boosted techniques to tag hadronically decaying $W$ and $Z$ gauge bosons, which strongly suppress the QCD dijet backgrounds. All three excesses emerge at around $2$ TeV in the invariant mass distribution formed by two $W$- or $Z$-tagged fat jets. 
The CMS collaboration also sees a moderate excess at the similar location in all hadronic channel \cite{Khachatryan:2014gha,Khachatryan:2014hpa}.
In response to the tantalizing experimental observations, several papers have already appeared taking this phenomenon as a new physics signature~\cite{Yepes:2015qwa, Bian:2015ota, Anchordoqui:2015uea, Chao:2015eea, Fukano:2015hga, 
Hisano:2015gna,Franzosi:2015zra,Cheung:2015nha,Dobrescu:2015qna,Aguilar-Saavedra:2015rna,Alves:2015mua,Gao:2015irw,Thamm:2015csa,Brehmer:2015cia,Cao:2015lia,Cacciapaglia:2015eea,Abe:2015jra,Allanach:2015hba,Abe:2015uaa,Carmona:2015xaa,Dobrescu:2015yba,Chiang:2015lqa,Cacciapaglia:2015nga,Fukano:2015uga,Sanz:2015zha,Omura:2015nwa,Chen:2015xql}.

A typical recipe for a new physics model to explain the above-mentioned excesses is the introduction of two new heavy states: a charged particle and a neutral particle. The former takes care of the $W^{\pm}Z$ channel while the latter does the other two channels. However, given the fact that a large fraction of events belong to all three channels, it may be a reasonable attempt to fit the data only with a single new heavy resonance.  As a matter of fact, Allanach, Gripaios and Sutherland recently  investigated the diboson resonances in this direction: they basically introduced a likelihood function for the true signal in the $W^+W^-$, $W^\pm Z$, and $ZZ$ channels and found that the maximum likelihood has zero events in the $W^\pm Z$ channel~\cite{Allanach:2015hba}. 
If this observation were true, the ATLAS data would indicate a single neutral bosonic resonating particle rather than two, which show up in all three channels due to detector effects and misidentification of $W^\pm$ and $Z$ bosons.  We also note that in the single particle interpretation coincidence of the resonances at $2$ TeV in the three channels can be naturally understood. Keeping this minimality and simplicity of the single particle interpretation, we further investigate the possible classification of neutral resonances by considering different spins  and CP states in an effective field theory approach including a set of operators for each case. 

Our philosophy is basically the bottom-up approach, invoking a minimal set of effective operators that may be responsible for the $W^+W^-$ and $ZZ$ signals. As no spin information is available, we extensively consider spin-0, spin-1, and spin-2 resonances. Symmetries of the relevant operators also induce potential signals in different final states, encouraging experimental collaborations to look into the related channels for consistency. 

In the following three sections, we examine scalar, vector, and tensor resonances in turn, focusing on viable parameter scans in conjunction with production cross sections and partial decay widths of the resonance at hand. In Section~\ref{sec:kinematics}, we briefly make comments on kinematic correlations among the final state particles to extract spin, CP states, and coupling information of the resonance of interest and the proposed interactions. Section~\ref{sec:discussion} is reserved for a summary.

%%%%%%%%%%%%%%%%%%%%%%%%%%%%%%%%%%%%%%%%%%%%%%%%%%%%%%%%%%%%
\section{Spin-0 resonances}   \label{sec:scalar}
%%%%%%%%%%%%%%%%%%%%%%%%%%%%%%%%%%%%%%%%%%%%%%%%%%%%%%%%%%%%

In our new physics interpretation, the resonance particle decays into two bosons so that the resonance itself should be a bosonic state with an integer spin.  In this section, we begin with considering a spin-0 resonance and study the effects of its CP states with corresponding effective operators.

%%%%%%%%%%%%%%%%%%%%%%%%%%%%%%%%%%%%%%%%%%%%%%%%%%%%%%%%%%%%
%\subsection*{CP-even spin-0 resonance} \label{sec:scalareven}
%%%%%%%%%%%%%%%%%%%%%%%%%%%%%%%%%%%%%%%%%%%%%%%%%%%%%%%%%%%%

A CP-even scalar resonance (henceforth denoted as $S$) in diboson channel could be well-parameterized by the following interaction Lagrangian:
\be
{\cal L}_s = -\frac{1}{\Lambda}S \Big(s_1 F^Y_{\mu\nu}{ F}^{Y\mu\nu} +s_2 F^W_{\mu\nu}{ F}^{W\mu\nu}+ s_3 G_{\mu\nu}^a{G}^{a\mu\nu}  \red{+\sum_f s_f m_f \bar{f}f} \Big),
\label{eq:L_even_s}
\ee
where $F_{\mu\nu}^Y$ and $F_{\mu\nu}^W$ denote the field strength tensors for usual U(1)$_Y$ and SU(2)$_W$ gauge bosons {\it before} the electroweak symmetry breaking while $G^a_{\mu\nu}$ denotes the SU(3)$_c$ gluon field strength tensor with the color index $a=1,2,\cdots 8$.\footnote{A scalar particle such as gravi-scalar or radion \cite{rad1, rad2, rad3} potentially provides diboson resonance and may have other signatures \cite{rad4, rad5, rad6}. However, we found that the width of 2 TeV radion is unacceptably big to account for the ATLAS anomaly.} The strengths of the above couplings are parametrized by $s_1$, $s_2$, and $s_3$, respectively \red{for gauge bosons and $s_f$ for fermions. }

\red{A tiny flavor non-diagonal interaction would lead un-acceptable flavor changing neutral current (FCNC) effects so that we naturally expect that the coefficients $s_f$ are negligibly small or strictly flavor diagonal. The first generation quarks, $u$ and $d$, could have the largest contribution to the production of the scalar at the LHC but they are suppressed by a small factor  $\sim m_f/\Lambda$.  
Furthermore, the coefficient $s_f$ can be forbidden by a global symmetry when the singlet scalar is promoted to a complex scalar field $T$ with ${\rm Re}(T)\equiv S$.  The couplings to the gauge bosons in the form of Eq.~({\ref{eq:L_even_s}}), however, are still obtained due to SM anomalies. A similar argument can be applied to the CP-odd spin-0 resonance. Considering all these, we would take the gluon fusion as the dominant production mechanism for the scalar resonance and neglect the production by diquark.}

Without loss of generality, we take $s_3=1$ by redefining $\Lambda$. The other coefficients, $s_1$ and $s_2$, for $U(1)_Y$ and $SU(2)_W$ gauge kinetic terms, are redefined as relative strengths to $s_3$. 
From the interactions in Eq.(\ref{eq:L_even_s}), we obtain the partial decay widths of $S$ into $\gamma\gamma$, $Z\gamma$, $ZZ$, $W^+W^-$, and $gg$ as 
\begin{eqnarray}
\begin{cases}
 \Gamma_S(\gamma\gamma)=\frac{|s_{\gamma\gamma}|^2 m_S^3}{4\pi \Lambda^2} \, ,
		& s_{\gamma\gamma}= s_1 \cos^2\theta_W+s_2 \sin^2\theta_W \, , \label{eq:twophotons} \\
\Gamma_S(ZZ) =  \frac{|s_{ZZ}|^2 m_S^3}{4\pi \Lambda^2}\sqrt{1-4x^S_Z}\left(1-4x^S_Z+6(x^S_Z)^2\right) \, ,
		&s_{ZZ}=s_2  \cos^2\theta_W+s_1 \sin^2\theta_W \, ,\\
\Gamma_S(Z\gamma )=\frac{|s_{Z\gamma}|^2 m_S^3}{8\pi \Lambda^2} \left(1-x^S_Z\right)^3 \, ,
		& s_{Z\gamma}= (s_2-s_1)\sin 2\theta_W \, ,\\
\Gamma_S(W^+W^-) = \frac{|s_{WW}|^2 m_S^3}{8\pi \Lambda^2}\sqrt{1-4x_W^S}\left(1-4x^S_W+6(x^S_W)^2\right),
		&s_{WW}=2 s_2 \, \\
\Gamma_S(gg)=\frac{2|s_{gg}|^2 m_S^3}{\pi \Lambda^2} \, , &s_{gg}=s_3 \, ,
\end{cases}
\end{eqnarray}
where $m_S$ and $\theta_W$ denote the mass of CP-even scalar S and the Weinberg angle. Here and henceforth, we define the mass squared ratio of a heavy SM boson $i$ ($Z$, $W$, or $h$) to a resonance $R$ as 
\bea
x^R_{i} \equiv \frac{m_i^2}{m_R^2}. \label{eq:ratio}
\eea 

Obviously, in this parametrization, $S$ is produced via gluon fusion followed by the decays into the above final states. Of potential experimental constraints, the two following conditions should be settled to be in the right ``ball park'' with respect to the recent ATLAS data:
\begin{itemize}\itemsep1pt \parskip0pt \parsep0pt
\item the total decay width should be within $\sim10$\% of the mass of the resonance \cite{Aad:2015owa}, 
\item the signal production cross section should be as sizable as order of several fb \cite{Allanach:2015hba}.
\end{itemize}
In general, the {\it single} production cross section of a narrow resonance is proportional to the total decay width of the decaying particle. Therefore, demanding a sizable production cross section with a (relatively) smaller total decay width is {\it not} a trivial task. We remark that as discussed in the literature, reported excesses in all three diboson final states ($W^+W^-$, $W^\pm Z$ and $ZZ$) are not independent of one another, and the data in one channel may be contaminated by the data in the other channels due to detector effects. As we mentioned in introduction, in Ref. \cite{Allanach:2015hba}, authors performed a general analysis of new physics interpretations of the recent ATLAS diboson excesses by computing a likelihood function for the true signal in the $W^+W^-$, $W^\pm Z$, and $ZZ$ channels. They found that the maximum likelihood has zero events in the $W^{\pm}Z$ channel, i.e., one could fit the data in all three channels with a {\it single} neutral resonance in the final state with 
$W^+W^-$ and $ZZ$. The likelihood is sufficiently flat and the required cross section (for 95\% C.L.) is in the following range \cite{Allanach:2015hba}:
\begin{equation}
{\mathcal O} (4-8)\; {\rm fb }~ \lesssim   \sigma \cdot BR(W^+W^-) + \sigma \cdot BR(ZZ) \lesssim  {\mathcal O}(20-24)\; {\rm fb } \, ,
\end{equation}
where $\sigma$ is the single production cross section of the resonance. 
For our analysis with the case of the CP-even scalar, we first fix the mass of the scalar resonance, $m_S$, to 2 TeV, and then compute the signal cross section, $\sigma ( p p \to S \to W^+W^- +ZZ )$, 
by varying three parameters, $\Lambda$, $s_1$ and $s_2$ ($s_3$=1). We find that in the majority of parameter space, the consistency (gauge invariance and Lorentz invariance) of the model predicts a large branching fraction into the diphoton final state. In particular, when two parameters have the same sign (i.e., $s_1 s_2 > 0$), diphoton rate ($\propto |s_{\gamma\gamma}|^2=|s_1\cos^2\theta_W +s_2 \sin^2 \theta_W|^2$) turns out to be too large so that the model is severely constrained by current data at the 8 TeV LHC~\cite{Aad:2015mna,CMS:2015cwa}.

%%%%%%%%%%%% 
\begin{figure}[t]
\begin{center}
\centerline{
\includegraphics[width=0.49\linewidth]{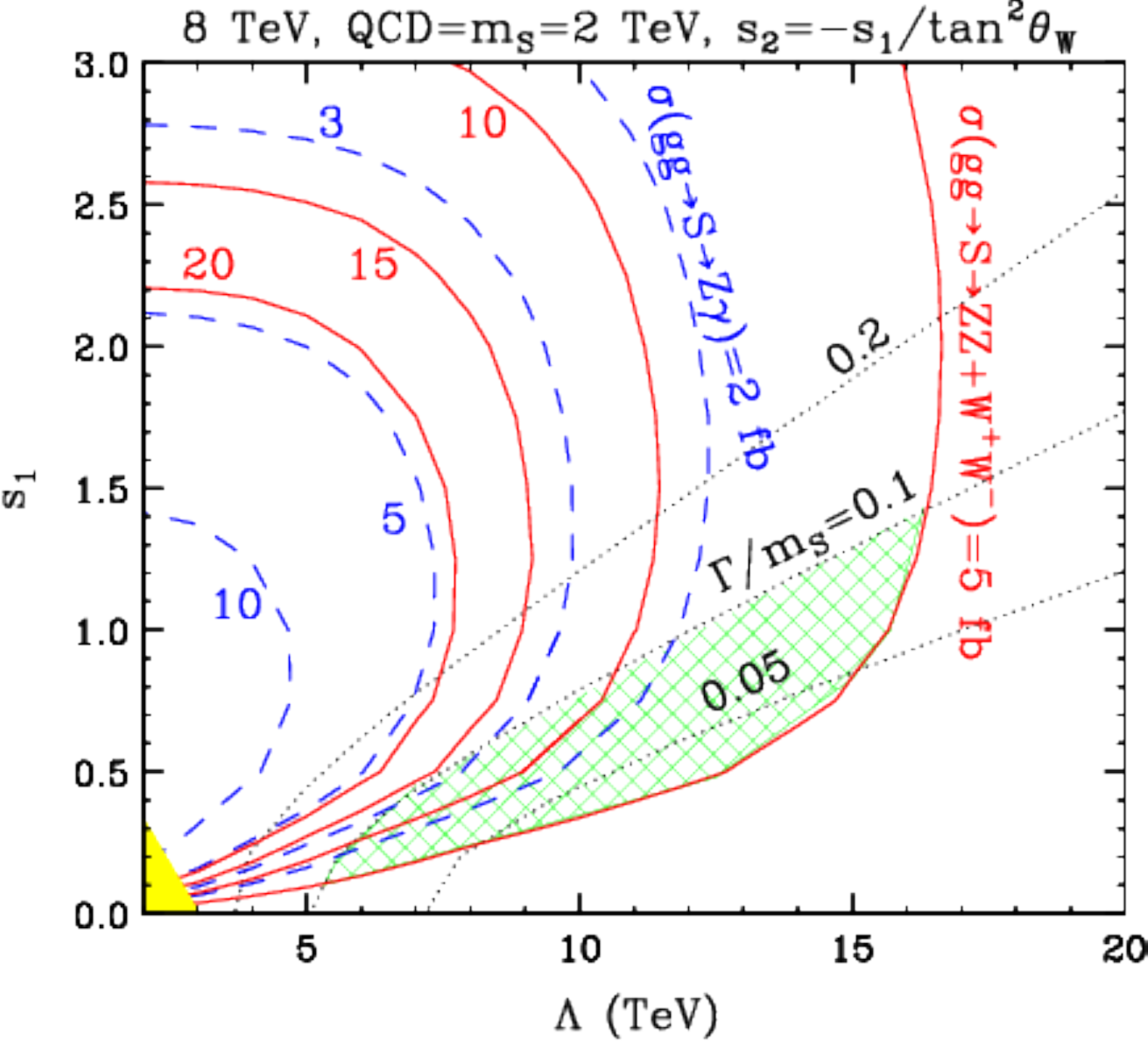} \hspace{0.2cm}
\includegraphics[width=0.50\linewidth]{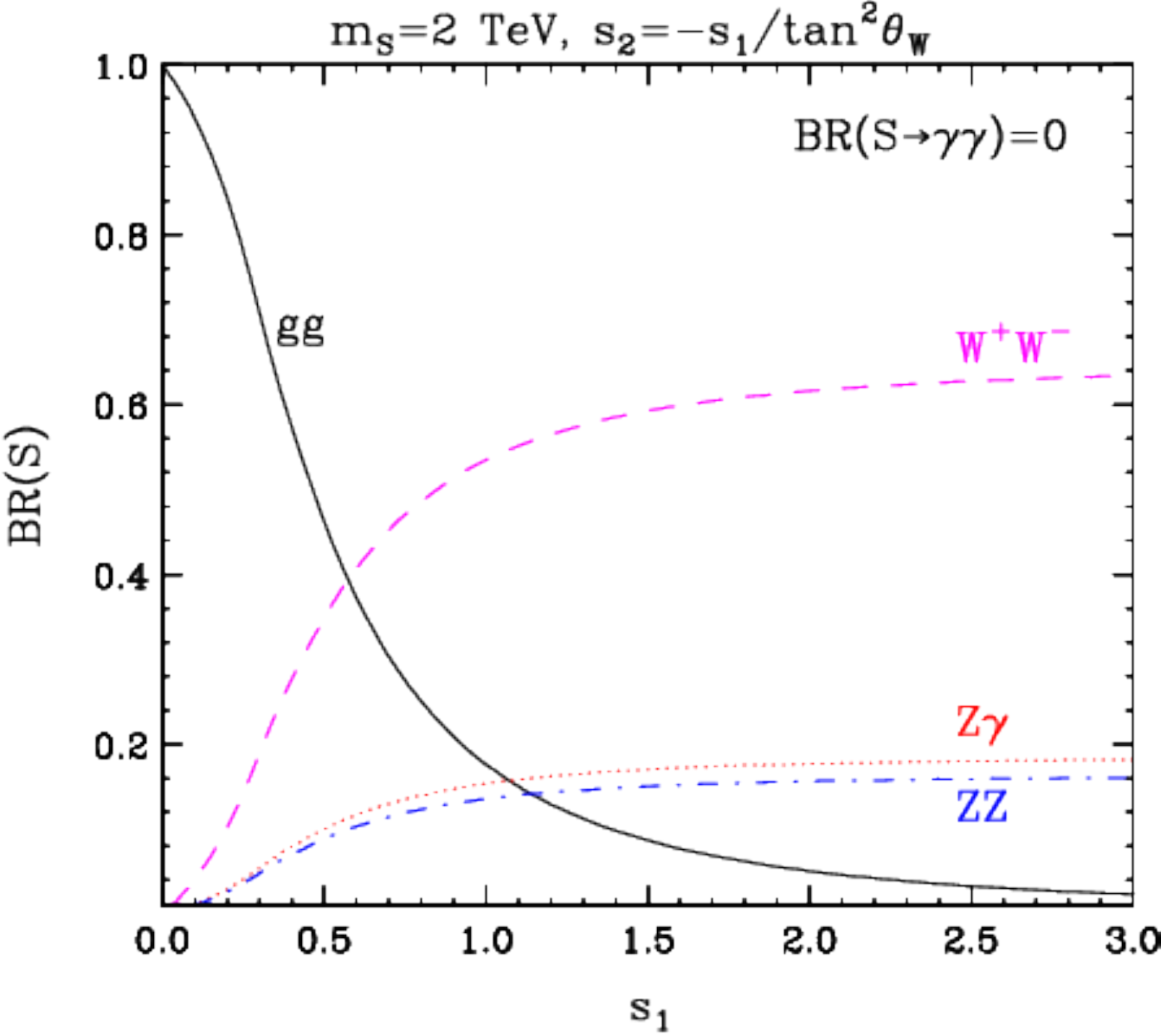}
}
\caption{
The left panel shows production cross sections (in fb) of the CP-even scalar resonance in the final states with $W^+W^- + ZZ$ (red solid curves) and $Z \gamma$ (blue dashed curves). 
The black dotted curves represent the constraint of the total decay width, $\Gamma/m_S$.
The dark yellow-shaded region is excluded by the dijet search and the light green-shaded region represents the allowed parameter space. 
The right panel shows branching fractions of the spin-0 resonance ($Z\gamma$, $ZZ$, $W^+W^-$, and $gg$ by red dotted, blue dot-dashed, magenta dashed, and black solid curves, respectively) as a function of $s_1$.}
\label{fig:scalareven}
\end{center}
\vspace*{-0.5cm}
\end{figure}
%%%%%%%%%%%%%

Interestingly enough, the opposite case with $s_1 s_2 < 0$ constraint provides a way to reduce the diphoton rate as clear from Eq.~\eqref{eq:twophotons}. 
Especially, the condition of $s_1 \approx - \tan^2\theta_W s_2$ gives $s_{\gamma\gamma}\approx 0$ thus vanishingly small diphoton final state, and the condition fixes the relative branching fractions as follows:
\begin{eqnarray}
BR(W^+W^-) : BR(ZZ) : BR(Z\gamma) : BR(gg) \nonumber \\
\approx \frac{|s_1|^2}{4 \tan^4\theta_W} : \frac{|s_1|^2 \cos^2 2\theta_W}{8\sin^4\theta_W } : \frac{|s_1|^2}{4 \tan^2\theta_W} : 1.
\end{eqnarray}
We take this relation for illustration and calculate the relevant cross sections in the two dimensional parameter space of $\Lambda$ vs. $s_1$, although other relations can be straightforwardly analyzed. For the relevant data analysis (and remaining analyses throughout this paper), we employ Monte Carlo event generators {\tt CalcHEP}~\cite{Belyaev:2012qa} and {\tt MadGraph5$\underline{\hspace{0.2cm} }$aMC@NLO}~\cite{Alwall:2011uj}. In Fig. \ref{fig:scalareven}, we show production cross sections (in fb) of the CP-even scalar resonance in the final states with $W^+W^-  + ZZ$ (red solid curves) and $Z \gamma$ (blue dashed curves). Contours of $\Gamma/m_S$ are shown by black-dotted curves. The corresponding branching fractions are shown in the right panel as a function of $s_1$. The dijet resonance searches provide constraints (at 95\% C.L.) on the parameters, which are shown by the dark yellow-shaded region~\cite{Khachatryan:2015sja,Aad:2014aqa}. 
Combining all constraints together, the allowed parameter space represented by the light green-shaded region might accommodate the diboson excesses. We remark that the exact relation of $s_1 = - \tan^2\theta_W s_2$ is not required, and any minor deviation from this relation would be easily allowed as long as the associated diphoton rate is below the current limit~\cite{Aad:2015mna,CMS:2015cwa}.

%%%%%%%%%%%%%%%%%%%%%%%%%%%%%%%%%%%%%%%%%%%%%%%%%%%%%%%%%%%%
%\subsection*{CP-odd spin-0 resonance}  \label{sec:scalarodd}
%%%%%%%%%%%%%%%%%%%%%%%%%%%%%%%%%%%%%%%%%%%%%%%%%%%%%%%%%%%%

Speaking of CP-odd spin-0 case, a pseudo-scalar or axion-like scalar (denoted as $A$) can couple to the SM gauge bosons through anomalies.
The gauge interactions are parametrized in a way similar to the CP-even scalar case with one of the field strength tensors replaced by a dual field strength tensor: 
\be
{\cal L}_a= -\frac{1}{\Lambda}A \Big(a_1 F^Y_{\mu\nu}{\tilde F}^{Y\mu\nu} +a_2 F^W_{\mu\nu}{\tilde F}^{W\mu\nu}+a_3 G_{\mu\nu}{\tilde G}^{\mu\nu}  \Big) \, ,
\ee
where the dual field strength tensors are defined as, for example, ${\tilde F}^Y_{\mu\nu}\equiv \frac{1}{2}\epsilon_{\mu\nu\rho\sigma} F^{Y\rho\sigma}$, and prefactors $a_1$, $a_2$, and $a_3$ denote the coupling constants which can be determined by anomalies for a global symmetry. 
For instance, $a_i/\Lambda=c_i \alpha_i/(8\pi f_A)\;(i=1,2,3)$ with $f_A$ being the breaking scale of a global $U(1)$ and $c_i=\sum_\alpha q_{\alpha} \ell_{G_i}(r_\alpha)$ where $q_\alpha$ is the global $U(1)$ charge of a heavy fermion and $\ell_{G_i}(r_\alpha)$ is the Dynkin index for a representation $r_\alpha$ under the SM gauge group $G_i$ \cite{AMDM,interplay}.

\begin{figure}[t!]
\begin{center}
\centerline{
\includegraphics[width=0.51\linewidth]{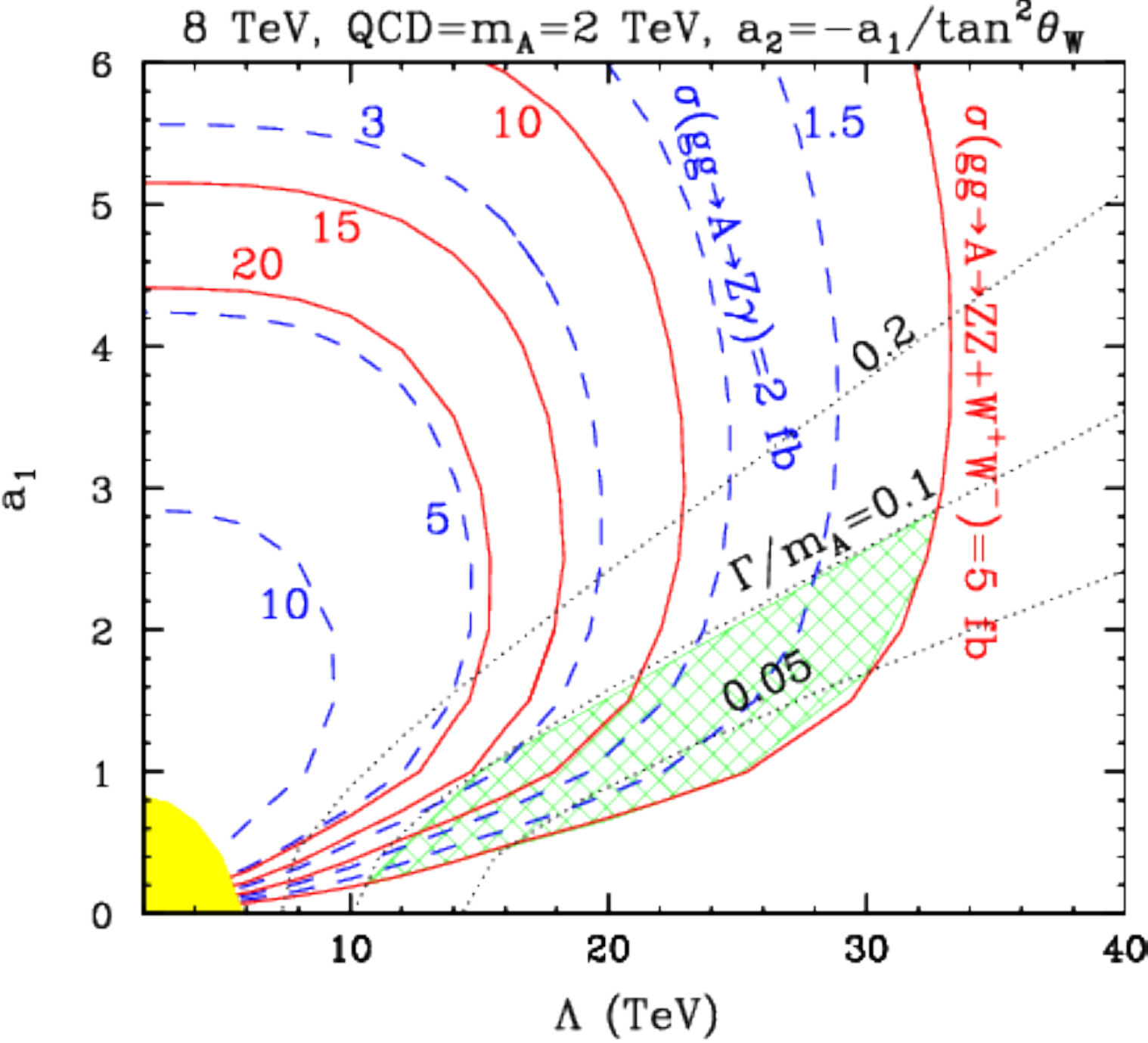}}
\caption{Same as in Fig. \ref{fig:scalareven} but for CP-odd scalar resonance. 
Branching fractions are similar to those for the CP-even scalar, replacing $s_i$ by $a_i$.}
\label{fig:scalarodd}
\end{center}
\vspace*{-0.5cm}
\end{figure}

The total decay width of the pseudo-scalar resonance~\cite{AMDM} is given by the sum of partial decay widths into $\gamma\gamma$, $Z\gamma$, $ZZ$, $W^+W^-$, and $gg$:  
\bea
\begin{cases}
 \Gamma_A(\gamma\gamma)=\frac{m^3_A}{4\pi \Lambda^2} |c_{\gamma\gamma}|^2, 
	& c_{\gamma\gamma}= a_1 \cos^2\theta_W+a_2 \sin^2\theta_W, \\
\Gamma_A(Z\gamma)= \frac{m^3_A}{8\pi \Lambda^2}|c_{Z \gamma }|^2\Big(1-x^A_Z\Big)^3, 
	&c_{Z\gamma}= (a_2 -a_1)\sin(2\theta_W), \\
\Gamma_A(ZZ)= \frac{m^3_A}{4\pi \Lambda^2} |c_{ZZ}|^2\Big(1-4x^A_Z\Big)^{3/2}, 
	&c_{ZZ}=a_2  \cos^2\theta_W+a_1 \sin^2\theta_W,\\
\Gamma_A(W^+W^-)=  \frac{m^3_A}{8\pi \Lambda^2}|c_{WW}|^2\Big(1-4x^A_W\Big)^{3/2}, 
	&c_{\tiny{WW}}= 2 a_2 \, , \\
\Gamma_A(gg)=\frac{2m^3_A}{\pi\Lambda^2} |c_{gg}|^2 \, , 
        &c_{gg}=a_3 \, ,
\label{GammAx}
\end{cases}
\eea
where $x^A_i$ is defined in Eq.~(\ref{eq:ratio}). The case with the CP-odd scalar has similarities compared to the case with the CP-even scalar in the sense that the corresponding branching fractions are similar along with associated coefficients, and also we require $a_1 a_2 < 0$ to suppress the diphoton rate. Like the CP-even scalar case we simply choose $a_2 = -a_1/\tan^2\theta_W$, and demonstrate the resulting parameter scans in Fig.~\ref{fig:scalarodd}. We observe that all contours are similar to those in Fig.~\ref{fig:scalareven}, except for the scale of $\Lambda$ due to a larger cross section for the CP-odd scalar.

A couple of comments should be made here. Speaking of the unitarity bound for scalar resonances first, we observe that in both CP-even and CP-odd cases, the spin-0 resonance couples only to transverse modes of SM gauge bosons. Then, the unitarity cutoff can be just read from the coefficients of the effective operators, namely, of order ${\rm max}(\Lambda/s_i)$ and ${\rm max}(\Lambda/a_i)$ in CP-even and -odd cases, respectively, by power counting. Thus, as shown in Fig. \ref{fig:scalareven}  and Fig. \ref{fig:scalarodd}, the unitarity cutoff is $\gtrsim \Lambda\sim 10\,{\rm TeV}$, which is consistent with the effective interactions with a TeV-scale resonance. Second, we find that in both CP-even and CP-odd cases, the $Z\gamma$ production cross section is about 1-3 fb at the 8 TeV in the allowed parameter space. The current experimental data tells that the (95\% C.L.) upper bound on the $Z\gamma$ production in the dilepton channel is given up to the resonance mass of 1.6 TeV while the higher mass reach is limited by statistics~\cite{Aad:2014fha}. Nevertheless, we expect that the corresponding limit for the 2 TeV resonance would be comparable to the result at the resonance mass of 1.6 TeV in $Z\gamma$ searches or below the existing limit. Therefore, $\sigma (p p \to Z\gamma) = {\cal O}(1)$ fb is still allowed for the 2 TeV and this channel would rather provide an interesting consistency check for $ZZ$ and $W^+W^-$ excesses. As shown in the right panel of Fig. \ref{fig:scalareven}, $BR(Z\gamma)$ is comparable to $BR(ZZ)$, and 
one cannot turn off $BR(Z\gamma)$, as it would also eliminate the signal.
In other words, if diboson excesses turned out to be the real signal with a CP-even or -odd scalar, observation of an excess in the $Z\gamma$ channel would corroborate the case. 
%
%%%%%%%%%%%% 
%%%%%%%%%%%%% 

%%%%%%%%%%%%%%%%%%%%%%%%%%%%%%%%%%%%%%%%%%%%%%%%%%%%%%%%%%%%
\section{Spin-1 resonances}  \label{sec:vector}
%%%%%%%%%%%%%%%%%%%%%%%%%%%%%%%%%%%%%%%%%%%%%%%%%%%%%%%%%%%%

For spin-1 resonances, we consider an extra $U(1)_X$ gauge symmetry that is realized by the Stueckelberg mechanism. Then, the would-be Goldstone boson $a_X$ ensures the gauge invariance of the effective action.

First, imposing the SM gauge symmetry and $U(1)_X$, we have the  dimension-4 interaction Lagrangian between the $U(1)_X$ gauge boson and the quarks and/or gauge bosons in the SM  given as follows:
\be
{\cal L}_{D4}=-g_XZ'_\mu {\bar q}\gamma^\mu(c_L P_L+c_R P_R)q -\frac{1}{2}\epsilon F^Y_{\mu\nu} F^{X\mu\nu}-\Big(i\eta D^\mu a_X\,(H^\dagger D_\mu H)+{\rm c.c.}\Big) \, , \label{dim4}
\ee
where the covariant derivative is defined as $D_\mu a_X \equiv \partial_\mu a_X -g_X Z'_\mu$ with $a_X$ being the Stueckelberg axion, $g_X$ is the $Z'$ gauge coupling, and $c_R=c_L$ ($c_R=-c_L$) for CP-even (CP-odd) $Z'$.  These dimension-4 interactions correspond to diquark couplings, gauge kinetic mixing and mass mixing in order.

 \red{We keep the dimension-4 diquark coupling to $Z'$ in Eq.~(\ref{dim4}) as a production mechanism, while the lepton couplings are suppressed as in leptophobic $Z'$ models \cite{leptophobic}. }
\red{The gauge kinetic mixing with $\epsilon \neq 0$ leads to $\Gamma(Z'\rightarrow Zh)=\Gamma(Z'\rightarrow WW)$ due to the SM gauge symmetry.  As the $Zh$ channel is strongly constrained by the LHC bound, $\sigma(pp\rightarrow Z')\times {\rm BR}(Z'\rightarrow Zh)\lesssim 7\,{\rm fb}$ \cite{Zh-bound}, which is significantly lower than  the required value, $\simeq 10\,{\rm fb}$, for explaining the ALTAS diboson excesses~\cite{Hisano:2015gna}.  Moreover, no $ZZ$ decay is induced from the dimension-4 operators. Therefore, we do not consider the possibility of a sizable mass mixing with $Z'$ any more taking $\epsilon \ll 1$. Instead, we consider novel effective interactions for $Z'$ containing the $ZZ$ decay mode, coming from dimension-6 operators, as will be discussed below.  \blue{The last term with real $\eta$ in Eq.~(\ref{dim4}), which is a CP-even interaction, should be highly suppressed, because of potential $Z'$ decays into $Z_Lh$ or $W_LW_L$.  If $\eta$ is purely imaginary, namely, the last term  in Eq.~(\ref{dim4}) is equivalent to a CP-odd operator $(\partial_\mu D^\mu a_X) H^\dagger H$ up to a total derivative, thus leading to $(\partial^\mu Z'_\mu)H^\dagger H$, but a vanishing on-shell decay amplitude squared for $Z'\rightarrow H^\dagger H$.} }
%induces $Z' \to 2H$. Either cases are acceptable so that we would take $\eta\ll 1$ below.}

%\red{The last term with real $\eta$ \footnote{We note that when $\eta$ is pure imaginary, the last term in eq.~(\ref{dim4}) is equivalent to a dimension-4 operator, $(\partial_\mu D^\mu a_X) H^\dagger H$ up to a total derivative. This interaction becomes $(\partial^\mu Z'_\mu)H^\dagger H$, which may introduce a bi-Higgs decay mode for $Z'$.}  reflects a charge shift of the SM Higgs doublet under the $U(1)'$ so $Z'$ could decay into $Z_L h$ or $W^+_L W^-_L$ with equal branching fractions. But, in the effective field theory with the SM and $Z'$ only, the charge shift under the $U(1)'$ must be universal for all the SM fermions with nonzero Yukawa couplings. Therefore, after the SM hypercharge is redefined, the third term becomes unphysical. }

It is noteworthy that the interactions of a vector isospin triplet $W'$ to the SM electroweak gauge bosons can be introduced by a similar dimension-4 operator in the effective theory such as $W^{\prime a}_\mu H^\dagger \sigma^a D^\mu H$ \cite{Allanach:2015hba}, which mixes  the extra gauge boson with the SM massive gauge bosons. In this case, the ATLAS diboson excesses can be explained by the $W^{\pm}Z$ channel, provided that the charged spin-1 resonance is produced via quark annihilation at the LHC~\cite{Allanach:2015hba}. 
In our work, we do not investigate the potential of the charged resonance as mentioned earlier because the dibosonic decays of a neutral resonance suffice to explain the ATLAS diboson excesses within current experimental errors.

Moving onto higher dimensional operators, we enumerate CP-even dimension-6 operators as follows~\cite{Dudas:2009uq}:
\bea
{\cal L}_{D6}&=&\frac{a_1}{\Lambda^2}\,D^\mu a_X[i (D^\nu H)^\dagger {\tilde F}^Y_{\mu\nu} H+{\rm c.c.}]
+\frac{a_2}{\Lambda^2}\,D^\mu a_X [(D^\nu H)^\dagger {F}^Y_{\mu\nu} H +{\rm c.c.}] \nonumber \\
&&+\frac{a_3}{\Lambda^2}\,D^\mu a_X[i (D^\nu H)^\dagger {\tilde F}^W_{\mu\nu} H+{\rm c.c.}]
+\frac{a_4}{\Lambda^2}\,D^\mu a_X [(D^\nu H)^\dagger { F}^W_{\mu\nu} H +{\rm c.c.}] \nonumber \\
&&+\frac{1}{\Lambda^2}\, \partial^\mu D_\mu a_X \Big(b_1 F^Y_{\rho\sigma}{\tilde F}^{Y\rho \sigma} +b_2F^W_{\rho\sigma}{\tilde F}^{W\rho \sigma} + b_3G_{\rho\sigma}{\tilde G}^{\rho \sigma}  \Big) \, ,
\label{d6even}
\eea
where $\Lambda$ is of order the mass of extra heavy fermions, and $a_i\;(i=1,2,3,4)$ and $b_i\;(i=1,2,3)$ parametrize the coupling strengths. The CP-odd counterparts of dimension-6 interactions are
\bea
{\tilde{\cal L}}_{D6}&=& \frac{{\tilde a}_1}{\Lambda^2}\,D^\mu a_X [i(D^\nu H)^\dagger { F}^Y_{\mu\nu} H+{\rm c.c.}]
+\frac{{\tilde a}_2}{\Lambda^2}\,D^\mu a_X [(D^\nu H)^\dagger {\tilde F}^Y_{\mu\nu} H +{\rm c.c.}] \nonumber \\
&&+\frac{{\tilde a}_3}{\Lambda^2}\, D^\mu a_X [i(D^\nu H)^\dagger { F}^W_{\mu\nu} H +{\rm c.c.}]
+\frac{{\tilde a}_4}{\Lambda^2}\,D^\mu a_X [(D^\nu H)^\dagger {\tilde F}^W_{\mu\nu} H +{\rm c.c.}] \nonumber \\
&&+\frac{1}{\Lambda^2} \,\partial^\mu D_\mu a_X \Big({\tilde b}_1 F^Y_{\rho\sigma}{F}^{Y\rho \sigma} +{\tilde b}_2F^W_{\rho\sigma}{F}^{W\rho \sigma} + {\tilde b}_3 G_{\rho\sigma}{G}^{\rho \sigma}  \Big),
\label{d6odd}
\eea
where $\tilde{a}_i\;(i=1,2,3,4)$ and $\tilde{b}_i\;(i=1,2,3)$ parametrize the coupling strengths. We comment on the dimension-6 operators composed of one field strength tensor for $Z'$ and two field strength tensors for the SM gauge bosons: ${\rm Tr}(F^{X\lambda}_\mu F_{\lambda\nu} {\tilde F}^{\nu\mu})$ for CP-even operators and ${\rm Tr}(F^{X\lambda}_\mu F_{\lambda\nu} {F}^{\nu\mu})$ for CP-odd operators with $F_{\mu\nu}=F^Y_{\mu\nu}, F^W_{\mu\nu}, G_{\mu\nu}$. First of all, the CP-odd operators can be rewritten as $F^{X\nu}_\mu F_{\nu\lambda} F^{\lambda\mu}=F^X_{\mu\nu} F_{\lambda\nu} F^{\mu\lambda}$, which is the same as $F^X_{\nu\mu} F_{\lambda\nu} F^{\mu\lambda}=-F^X_{\mu\nu} F_{\lambda\nu}F^{\mu\lambda}$, and as a result, we get $F^{X\nu}_\mu F_{\nu\lambda} F^{\lambda\mu}=0$. Likewise, the CP-even operators can be also rewritten as $F^{X\nu}_\mu F_{\nu\lambda} {\tilde F}^{\lambda\mu}=F^X_{\mu\nu} F_{\lambda\nu} {\tilde F}^{\mu\lambda}$.  Then, using the identity of $F_{\lambda\mu} {\tilde F}^{\nu\lambda}=-\frac{1}{4}\delta^\nu_\mu\, F_{\alpha\beta}{\tilde F}^{\alpha\beta}$, we get $F^{X\nu}_\mu F_{\nu\lambda} {\tilde F}^{\lambda\mu}=-\frac{1}{4}F^{X\mu}_\mu F_{\alpha\beta}{\tilde F}^{\alpha\beta}=0$.
Therefore, the dimension-6 operators composed of gauge field strength tensors only are identically zero so that we do not consider them in our analysis.

Given the above observations, the $Z'$ gauge boson decays only by symmetry breaking terms given in  ${\cal L}_{D6}$ or ${\tilde{\cal L}}_{D6}$. 
When it comes to the production modes for the spin-1 resonance, we henceforth assume that it is produced by diquark couplings and ignore the gauge kinetic mixing and mass mixing. The effective cubic interactions for $Z'$ coming from ${\cal L}_{D6}$ and ${\tilde{\cal L}}_{D6}$ are obtained as shown below:
\bea
{\cal L}_{\rm CP-even}&=&\frac{v}{\Lambda^2}\Big( a_1 m_Z  Z^\nu Z^{\prime\mu}  {\tilde F}^Y_{\mu\nu}  +a_2 \partial^\nu h\, Z^{\prime \mu}F^Y_{\mu\nu} \Big) \nonumber \\
&&-\frac{v}{\Lambda^2} \Big(\frac{1}{2} a_3 m_Z\epsilon^{\mu\nu\rho\sigma}  Z_\nu Z'_\mu (\partial_\rho W^3_\sigma-\partial_\sigma W^3_\rho )  +a_4 \partial^\nu h \, Z^{\prime \mu}(\partial_\mu W^3_\nu-\partial_\nu W^3_\mu)  \Big) \nonumber \\
&&+\frac{m_W v}{\Lambda^2}Z'_\mu \Big(-\frac{1}{2} a_3  \epsilon^{\mu\nu\rho\sigma} W^-_\nu (\partial_\rho W^+_\sigma- \partial_\sigma W^+_\rho)+ia_4 W^{-\nu} (\partial_\mu W^{+}_\nu-\partial_\nu W^{+}_\mu)+{\rm c.c.}  \Big)  \nonumber \\
&&+ \frac{1}{\Lambda^2} \partial^\mu Z'_\mu \Big(b_1 F^Y_{\rho\sigma}{\tilde F}^{Y\rho \sigma} +b_2F^W_{\rho\sigma}{\tilde F}^{W\rho \sigma} + b_3G_{\rho\sigma}{\tilde G}^{\rho \sigma}  \Big) \, ,
\label{cpeven}
\eea
%and
\bea
{\cal L}_{\rm CP-odd}&=&\frac{v}{\Lambda^2}\Big({\tilde a}_1 m_Z  Z^\nu Z^{\prime\mu}  {F}^Y_{\mu\nu}  +{\tilde a}_2 \partial^\nu h\, Z^{\prime\mu}  {\tilde F}^Y_{\mu\nu}\Big) \nonumber \\
&&-\frac{v}{\Lambda^2} \Big( {\tilde a}_3 m_Z Z^\nu Z^{\prime \mu} (\partial_\mu W^3_\nu-\partial_\nu W^3_\mu)  + \frac{1}{2} {\tilde a}_4 \epsilon^{\mu\nu\rho\sigma}  \partial_\nu h \,Z^{\prime}_\mu (\partial_\rho W^3_\sigma-\partial_\sigma W^3_\rho) \Big)  \nonumber \\
&&+\frac{m_W v}{\Lambda^2}Z'_\mu \Big(-{\tilde a}_3W^-_{\nu} (\partial^\mu W^{+\nu}-\partial^\nu W^{+\mu})+\frac{1}{2} i{\tilde a}_4  \epsilon^{\mu\nu\rho\sigma} W^-_\nu (\partial_\rho W^+_\sigma- \partial_\sigma W^+_\rho)    +{\rm c.c.} \Big) \nonumber \\
&&+\frac{1}{\Lambda^2} \partial^\mu Z'_\mu \Big({\tilde b}_1 F^Y_{\rho\sigma}{F}^{Y\rho \sigma} +{\tilde b}_2F^W_{\rho\sigma}{F}^{W\rho \sigma} + {\tilde b}_3 G_{\rho\sigma}{G}^{\rho \sigma}  \Big)  \,, \label{cpodd}
\eea
where the $U(1)_X$ gauge coupling is absorbed into $a_1$ and ${\tilde a}_1$, and so on.

After the electroweak symmetry breaking (EWSB) and dropping the terms with the divergence of $Z'$, the effective CP-even interactions for $Z'$ are
\bea
{\cal L}_{\rm CP-even}&=& \kappa_1  \epsilon^{\mu\nu\rho\sigma} Z^\prime_ \mu Z_\nu F_{\rho\sigma}+ {\hat \kappa}_1 \epsilon^{\mu\nu\rho\sigma} Z'_\mu Z_\nu (\partial_\rho Z_\sigma-\partial_\sigma Z_\rho) \nonumber \\
 &&+\Big(\kappa_2 \epsilon^{\mu\nu\rho\sigma} Z'_\mu W^-_\nu (\partial_\rho W^+_\sigma - \partial_\sigma W^+_\rho)+i{\hat\kappa}_2 Z^{\prime\mu} W^{-\nu} (\partial_\mu W^{+}_\nu - \partial_\nu W^+_\mu)   +{\rm c.c.}\Big) \nonumber \\
&&+\frac{\kappa_3}{\Lambda} \, Z^{\prime\mu} \partial^\nu  h\, F_{\mu\nu}+ \frac{{\hat\kappa}_3}{\Lambda} \, Z^{\prime\mu} \partial^\nu  h \, (\partial_\mu Z_\nu-\partial_\nu Z_\mu) \,  \label{d6eff1}
\eea
where $F_{\mu\nu}$ is the photon field strength tensor and $\kappa_2, {\hat\kappa}_2$ are related to other parameters by gauge invariance as
\bea
\kappa_2 = \frac{m_W}{m_Z}\Big(\kappa_1 \sin\theta_W +{\hat\kappa}_1 \cos\theta_W\Big),\quad
{\hat\kappa}_2=-\frac{m_W}{\Lambda}(\kappa_3 \sin\theta_W+{\hat\kappa}_3\cos\theta_W ).
\eea
We note that the effective triple gauge interactions with $Z'$ in the above effective Lagrangian are the generalized Chern-Simons terms that are generated by extra heavy fermions \cite{Dudas:2009uq,interplay}.
Using the effective action above, we obtain the partial decay rates of the spin-1 resonance~\cite{interplay} into $Z\gamma$, $ZZ$, $W^+W^-$, $h\gamma$, $hZ$, and $q\bar{q}$, respectively as:
%\be
%\Gamma_{Z'}=\Gamma_{Z'}(Z\gamma)+\Gamma_{Z'}(ZZ)
%+\Gamma_{Z'}(WW)+\Gamma_{Z'}(h\gamma)+\Gamma_{Z'}(hZ)+\Gamma_{Z'}(q{\bar q}) \, ,
%\ee
%where  
\bea
\left\{
\begin{array}{l}
\Gamma_{Z'}(Z\gamma)=\frac{\kappa^2_1 m^3_{Z'}}{24\pi m_Z^2}\,\left(1-x_Z^{Z'}\right)^3 \Big(1+x_Z^{Z'}\Big), \\ %  \label{ZpZg} \\
\Gamma_{Z'}(ZZ)=\frac{{\hat\kappa}^2_1m^3_{Z'}}{24\pi m^2_Z} \, \Big(1-4x_Z^{Z'}\Big)^{5/2}, \\% \label{ZpZZ}  \\
\Gamma_{Z'}(W^+W^-)= \frac{ m^3_{Z'} \left (1-4x_W^{Z'}\right)^{3/2}}{48\pi m^2_W} 
\left [  4 \kappa_2^2   \left (1-4x_W^{Z'}\right) + \hat\kappa_2^2  \left (1+3x_W^{Z'}\right)     \right ], \\ % \label{ZpWW}  \\
\Gamma_{Z'}(h\gamma)= \frac{\kappa^2_3 m^3_{Z'}}{96\pi \Lambda^2} \Big(1-x_h^{Z'}\Big)^3,  \\
\Gamma_{Z'}(hZ)= \frac{{\hat\kappa}^2_3m^3_{Z'}}{192\pi \Lambda^2} \Big(1-(\sqrt{x_h^{Z'}}+\sqrt{x_Z^{Z'}})^2\Big)^{1/2} \Big(1-(\sqrt{x_h^{Z'}}-\sqrt{x_Z^{Z'}})^2\Big)^{1/2}   \\
\hspace{1.5cm}\times \left (  2 + x_Z^{Z'}  ( x_h^{Z'} - x_Z^{Z'} )^2  + 2 x_h^{Z'} (x_h^{Z'} + 3 x_Z^{Z'})- (4 x_h^{Z'} +3 x_Z^{Z'}) \right )
,   \\
\Gamma_{Z'}( q{\bar q})= \frac{g^2_Xm_{Z'}}{4 \pi } (1+2 x_q^{Z'}) \Big(1-4x_q^{Z'}\Big)^{1/2},
\end{array} \right.
\eea
where $x_i^{Z'}$ is defined in Eq.~(\ref{eq:ratio}). On the other hand, the effective CP-odd interactions for $Z'$ become
\bea
{\cal L}_{\rm CP-odd}&=& \alpha_1  Z'^\mu Z^\nu F_{\mu\nu}+ {\hat\alpha}_1  Z'^\mu Z^\nu (\partial_\mu Z_\nu -\partial_\nu Z_\mu) \nonumber \\
&&+\Big(\alpha_2  Z'^\mu W^{-\nu}( \partial_\mu W^+_\nu - \partial_\nu W^{+}_\mu) +i{\hat\alpha}_2  \epsilon^{\mu\nu\rho\sigma} Z'_\mu W^-_\nu (\partial_\rho W^+_\sigma - \partial_\sigma W^+_\rho)
 +{\rm c.c}\Big) \nonumber \\
&&+\frac{\alpha_3}{\Lambda} \,  \epsilon^{\mu\nu\rho\sigma} Z'_\mu \partial_\nu  h\, F_{\rho\sigma}
+\frac{{\hat\alpha}_3}{\Lambda} \,  \epsilon^{\mu\nu\rho\sigma} Z'_\mu \partial_\nu  h\, (\partial_\rho Z_\sigma-\partial_\sigma Z_\rho) \,  ,\label{d6eff2}
\eea
where $\alpha_2, {\hat\alpha}_2$ are related to other parameters by gauge invariance as
\bea
\alpha_2 = \frac{m_W}{m_Z}\Big(\alpha_1 \sin\theta_W+{\hat\alpha}_1 \cos\theta_W \Big), \quad
{\hat\alpha}_2 =- \frac{m_W}{\Lambda}\Big(\alpha_3 \sin\theta_W+{\hat\alpha}_3 \cos\theta_W \Big).
\eea
The partial decay widths of the CP-odd vector into $Z\gamma$, $ZZ$, $W^+W^-$, $h\gamma$, $hZ$, and $q\bar{q}$, respectively are given as follows: 
\begin{eqnarray}
\left\{
\begin{array}{l}
\Gamma_{Z'}(Z\gamma)= \frac{\alpha^2_1  m^3_{Z'}}{96\pi m_Z^2}\,\left(1-x_Z^{Z'}\right)^3 \Big(1+x_Z^{Z'}\Big)  \, ,  \\
\Gamma_{Z'}(ZZ)= \frac{ \hat\alpha_1^2 m^3_{Z'}}{96\pi m^2_Z} \, \Big(1-4x_Z^{Z'}\Big)^{3/2} \, , \\
\Gamma_{Z'}(W^+W^-)= 
 \frac{ m^3_{Z'} \sqrt{ 1-4x_W^{Z'}}} {48\pi m^2_W} 
\left [   \alpha_2^2   \left (1-4x_W^{Z'}\right) + 4 \hat\alpha_2^2  \left (1 + 2  x_W^{Z'}\right)     \right ],  \\
\Gamma_{Z'}(h\gamma)=   \frac{\alpha^2_3  m^3_{Z'}}{24\pi \Lambda^2} \Big(1-x_h^{Z'}\Big)^3 \, ,  \\
\Gamma_{Z'}(hZ)= \frac{  \hat\alpha_3^2  m^3_{Z'}}{24\pi \Lambda^2} \Big(1-(\sqrt{x_h^{Z'}}+\sqrt{x_Z^{Z'}})^2\Big)^{3/2} \Big(1-(\sqrt{x_h^{Z'}}-\sqrt{x_Z^{Z'}})^2\Big)^{3/2}  \, ,   \\
\Gamma_{Z'}( q{\bar q})= \frac{g^2_X m_{Z'}}{4 \pi } \Big(1-4x_q^{Z'}\Big)^{3/2} \, .
\end{array}\right.
\end{eqnarray}

As can be seen clearly from the gauge invariant higher dimensional operators in Eqs.~(\ref{d6even}) and (\ref{d6odd}) and can be checked from the effective gauge interactions in Eqs.~(\ref{d6eff1}) and (\ref{d6eff2}), we note that the unitarity cutoff of $\Lambda\sim 10\,{\rm TeV}$, implies that $\kappa_{1,2}, {\hat\kappa}_{1,2}, \alpha_{1,2}, {\hat\alpha}_{1,2}\lesssim {\cal O}(10^{-2})$ and  $\kappa_3,\alpha_3\lesssim {\cal O}(1)$.

\begin{figure}[t]
\begin{center}
\centerline{
\includegraphics[width=0.51\linewidth]{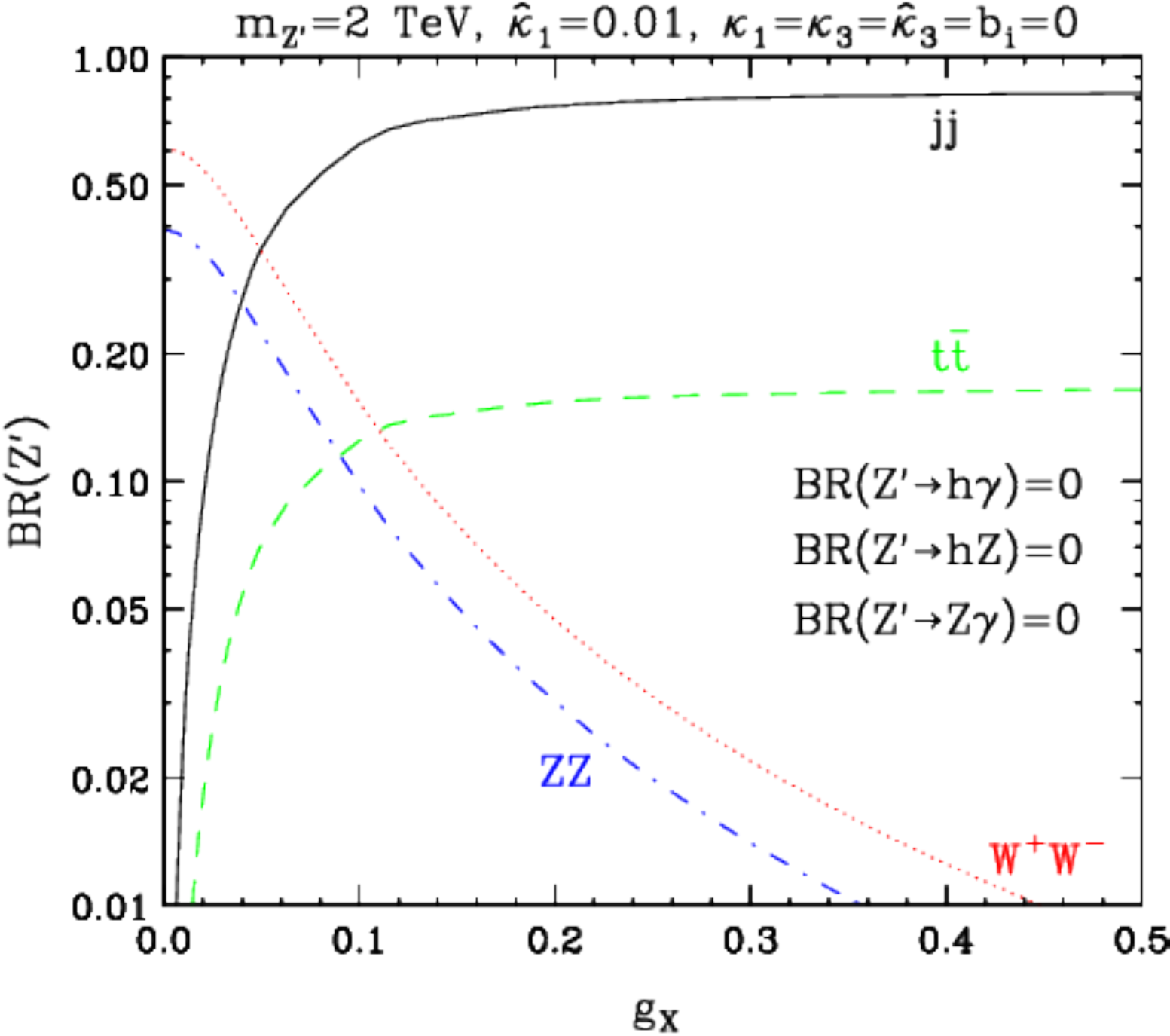} }
\caption{Branching fractions of the CP-even vector as a function of $g_X$ (diquark coupling) for a given set of 
parameters, $\hat\kappa_1 =0.01$ and $\kappa_1=\kappa_3=\hat\kappa_3=b_i =0$.
Very similar results are obtained for the CP-odd vector and the CP-odd tensor cases.
}
\label{fig:vectorBR}
\end{center}
\vspace*{-0.5cm}
\end{figure}

For a phenomenological study of the spin-1 resonance, we assume that the higher dimensional operators given in Eqs.~(\ref{cpeven}) and (\ref{cpodd}) come with pure imaginary coefficients, i.e. $a_2=a_4=0$ and ${\tilde a}_2={\tilde a}_4=0$. Then, we get 
\be
\kappa_3={\hat\kappa}_3=0, \quad {\tilde\kappa}_2=0, \quad \kappa_2=\frac{m_W}{m_Z}(\kappa_1\sin\theta_W +{\hat\kappa}_1 \cos\theta_W) \, ,
\ee
for CP-even interactions, and, similarly,   
\be
\alpha_3={\hat\alpha}_3=0, \quad {\hat\alpha}_2=0,\quad  \alpha_2=\frac{m_W}{m_Z}(\alpha_1\sin\theta_W +{\hat\alpha}_1 \cos\theta_W) \, ,
\ee
for CP-odd interactions. There are two free parameters for SM gauge boson couplings in each case, $\kappa_1,{\hat\kappa}_1$ and $\alpha_1, {\hat\alpha}_1$, respectively.
In this case, there are no $h\gamma$ or $hZ$ decay modes of the $Z'$ gauge boson while $Z\gamma, ZZ$ and $W^+W^-$ decay modes exist. Therefore,  the gauge invariance of the higher dimensional operators is crucial in correlating between different decay channels of the spin-1 resonance. Turning on small couplings to Higgs, we can maintain the diboson resonances as hinted by ATLAS and at the same time have a potential to discover or constrain the models with spin-1 resonance further by the decay mode into $h\gamma$ or $hZ$. 
Henceforth, in order to explain the ATLAS diboson excess from $W^+W^-$ and$ZZ$ decay modes, we focus on a simple parameter choice with $\kappa_3={\hat \kappa}_3=0$ for the CP-even and $\alpha_3={\hat \alpha}_3=0$ for the CP-odd. 
In this case, the ratio between $W^+W^-$ and $ZZ$ branching fractions remains constant, 
independent of the remaining parameters for both cases, i.e.,  
$\frac{BR( Z^\prime \to W^+W^-)}{R( Z^\prime \to Z Z) }\approx 1.56$.  
In Fig. \ref{fig:vectorBR}, we show branching fractions of the CP-even vector 
as a function of the diquark coupling ($g_X$) for above choice of parameters.
In addition, we have fixed $\hat\kappa_1 =0.01$ (considering the unitarity bound) to maximize the branching fraction into $W^+W^-$ and $ZZ$, as their partial decays widths are proportional to $\hat\kappa_1^2$ for 
$\kappa_1=\kappa_3=\hat\kappa_3=b_i =0$. 
Numerically very similar results are obtained for the CP-odd vector.
%
%%%%%%%%%%%% 
%%%%%%%%%%%%% 

\begin{figure}[t]
\begin{center}
\centerline{
\includegraphics[width=0.49\linewidth]{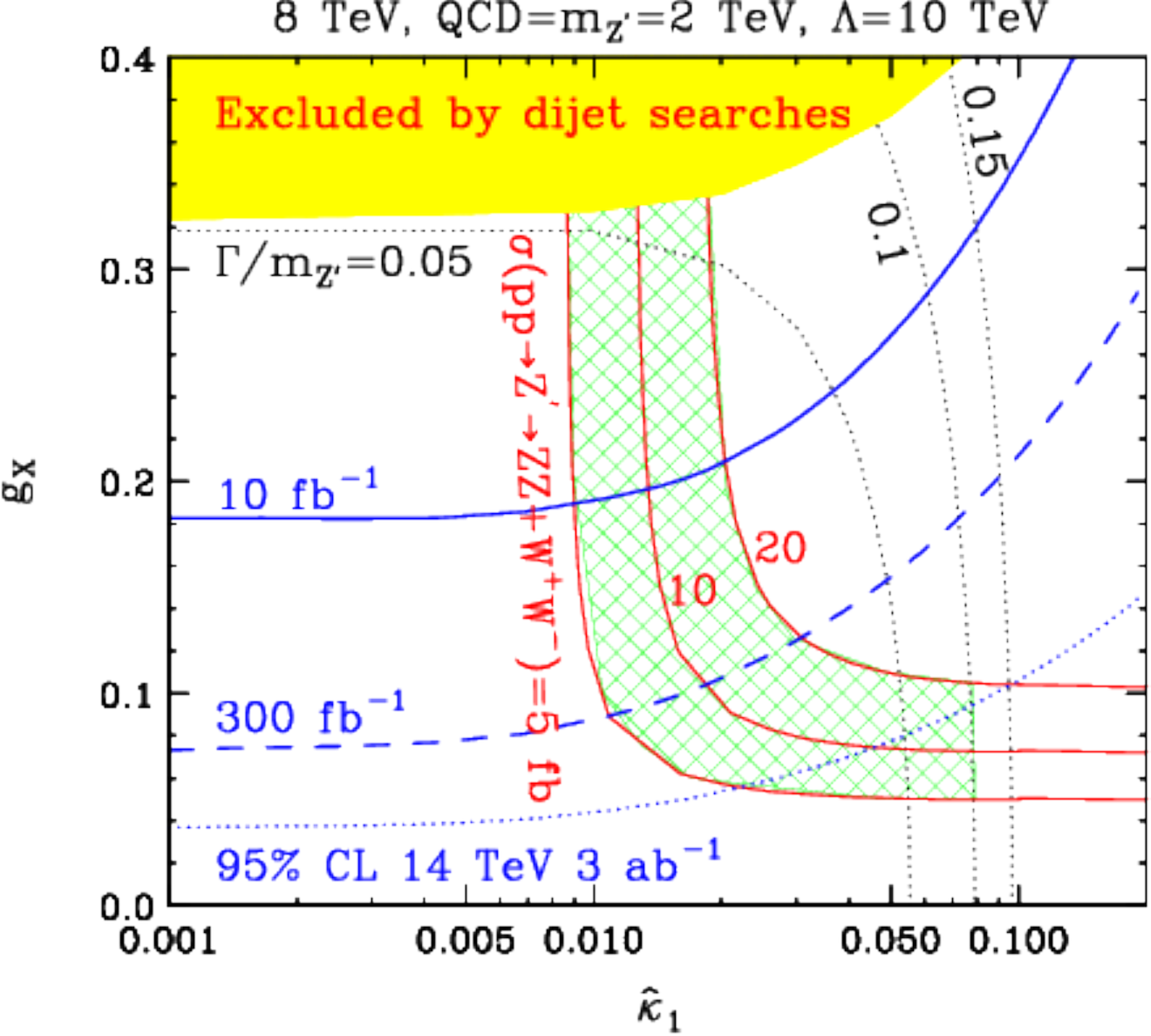} \hspace{0.05cm}
\includegraphics[width=0.49\linewidth]{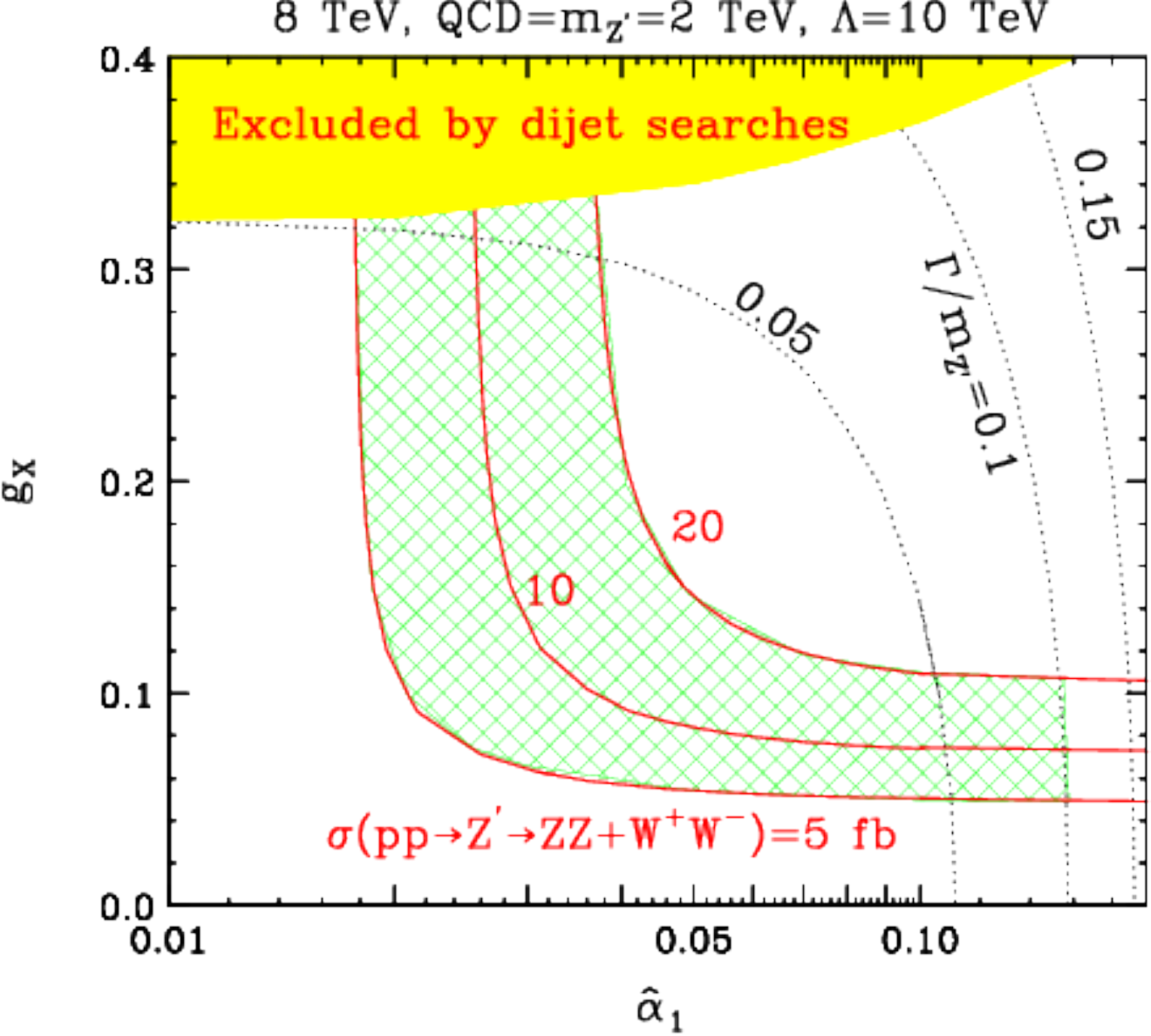} }
\caption{Production cross sections (in fb) of the CP-even (left panel) and the CP-odd (right panel) vector bosons in the $ZZ +W^+W^-$ final state.
The dark yellow-shaded region is disfavored by ATLAS dijet searches and the black dotted curves represent $\Gamma_{Z^\prime}/m_{Z^\prime} =0.15$, 0.1, and 0.05, respectively. The light green-shaded region represents the allowed space to fit the ATLAS diboson data.
The blue (solid, dashed, dotted) curves represent the projected 95\% C.L. exclusion contours for 14 TeV LHC with the corresponding luminosity.}
\label{fig:vector}
\end{center}
\vspace*{-0.5cm}
\end{figure}

For our numerical study, we set $c_L=c_R=1$ for the CP-even ($-c_L=c_R=1$ for the CP odd) and ignore the kinetic mixing and mass mixing.
We further set $b_i=0$ for the CP-even case ($\tilde b_i=0$ for the CP-odd case), leaving $\kappa_1$, $\hat\kappa_1$, $\kappa_3$, $\hat\kappa_3$, $\Lambda$ and $g_X$ for the CP-even, and 
$\alpha_1$, $\hat\alpha_1$, $\alpha_3$, $\hat\alpha_3$, $\Lambda$ and $g_X$ for the CP-odd, respectively, as relevant parameters.
Dependence on $\kappa_3$, $\hat\kappa_3$, $\alpha_3$ and $\hat\alpha_3$ are weak, and we set 
them to zero as mentioned above to make $\sigma ( h \gamma )$ and $\sigma( h Z )$ vanish.
Furthermore, we conservatively take $\kappa_1 = 0 = \alpha_1$, for which $\sigma (Z \gamma)$ also vanishes.
Turning on non-zero values of $\kappa_1$ and $\alpha_1$ always reduces the branching fractions of the diboson signal.
Finally, after setting $\Lambda=10$ TeV, 
we show in Fig. \ref{fig:vector} the production cross sections of the CP-even (left panel) and the CP-odd (right panel) vector bosons in the $ZZ+W^+ W^-$ final state (red solid curves).
As the resonance is produced by $pp$ collision, it can also decays to the dijet final state. 
The dark yellow-shaded area is disfavored by ATLAS dijet searches~\cite{Aad:2014aqa} and the black dotted curves represent $\Gamma_{Z^\prime}/m_{Z^\prime} =0.15$, 0.1,  and 0.05, respectively.  
 
The single production cross section itself is explicitly dependent on the coupling, $g_X$, only. 
However, the decay width changes depending on the rest of parameters, 
which affect the shape of the dijet cross section.
Our CP-even vector model is the same as one in discussed in Ref.  \cite{Yu:2013wta}, 
and we are able to use results there by simply rescaling couplings and branching fractions in our parameter space. 
The blue (solid, dashed, dotted) curves labelled by 10 fb$^{-1}$, 300 fb$^{-1}$, and 3 ab$^{-1}$ represent 
the projected 95\% C.L. exclusion contours for 14 TeV LHC, respectively. 
Unfortunately, this projected sensitivity is not available for other resonances, and it is not straightforward to recast the results from Ref. \cite{Yu:2013wta} due to different efficiencies. 

We note that as shown in the right panel of Fig. \ref{fig:vector}, the allowed parameter space requires 
$\hat \alpha_1 \gsim 0.02$, which is close to the unitarity limit. 
Finally, any reasonable deviation from the current choice of parameters would be easily allowed, 
as long as the corresponding limits can be avoided in the final states with $Z\gamma$, $hZ$, and $h\gamma $.
%%%%%%%%%%%% 
%%%%%%%%%%%%% 
%

%%%%%%%%%%%%%%%%%%%%%%%%%%%%%%%%%%%%%%%%%%%%%%%%%%%%%%%%%%%%
\section{Spin-2 resonances} \label{sec:tensor}
%%%%%%%%%%%%%%%%%%%%%%%%%%%%%%%%%%%%%%%%%%%%%%%%%%%%%%%%%%%%

The spin-2 resonance ${\cal G}_{\mu\nu}$ with mass $m_G$ couples to the SM particles as graviton does, that is, 
\be
{\cal L}^{\cal G}_{\rm CP-even}=\frac{1}{\Lambda} {\cal G}_{\mu\nu} T^{\mu\nu} \, ,
\ee
where $T_{\mu\nu}$ is the energy-momentum tensor.  We set the spin-2 resonance to couple to the energy-momentum tensor for each SM particle with an arbitrary coefficient, which is gauge invariant under the SM gauge groups.  
The energy-momentum tensor with CP-even interactions to the SM gauge bosons are
\be
T_{\mu\nu}=
c_1 F^Y_{\mu\lambda}F^{Y\lambda}\,_\nu+c_2 F^W_{\mu\lambda}F^{W\lambda}\,_\nu+c_3 G_{\mu\lambda}G^\lambda\,_\nu \, ,
\ee
where $c_1$, $c_2$, and $c_3$ are constant coefficients parametrizing the relevant coupling strengths. Here, we assumed that  the spin-2 resonance couples dominantly to the transverse modes of SM gauge bosons \cite{GMDM} while the terms proportional to the metric $g_{\mu\nu}$ in the energy-momentum tensor vanish under the traceless condition. 
For a heavy spin-2 resonance with $m_G\gg m_{W,Z}$, the gauge boson mass terms can be ignored, even if the spin-2 resonance couples to the longitudinal modes of gauge bosons as well~\cite{GMDM}. 

\begin{figure}[t!]
\begin{center}
\centerline{
\includegraphics[width=0.5\linewidth]{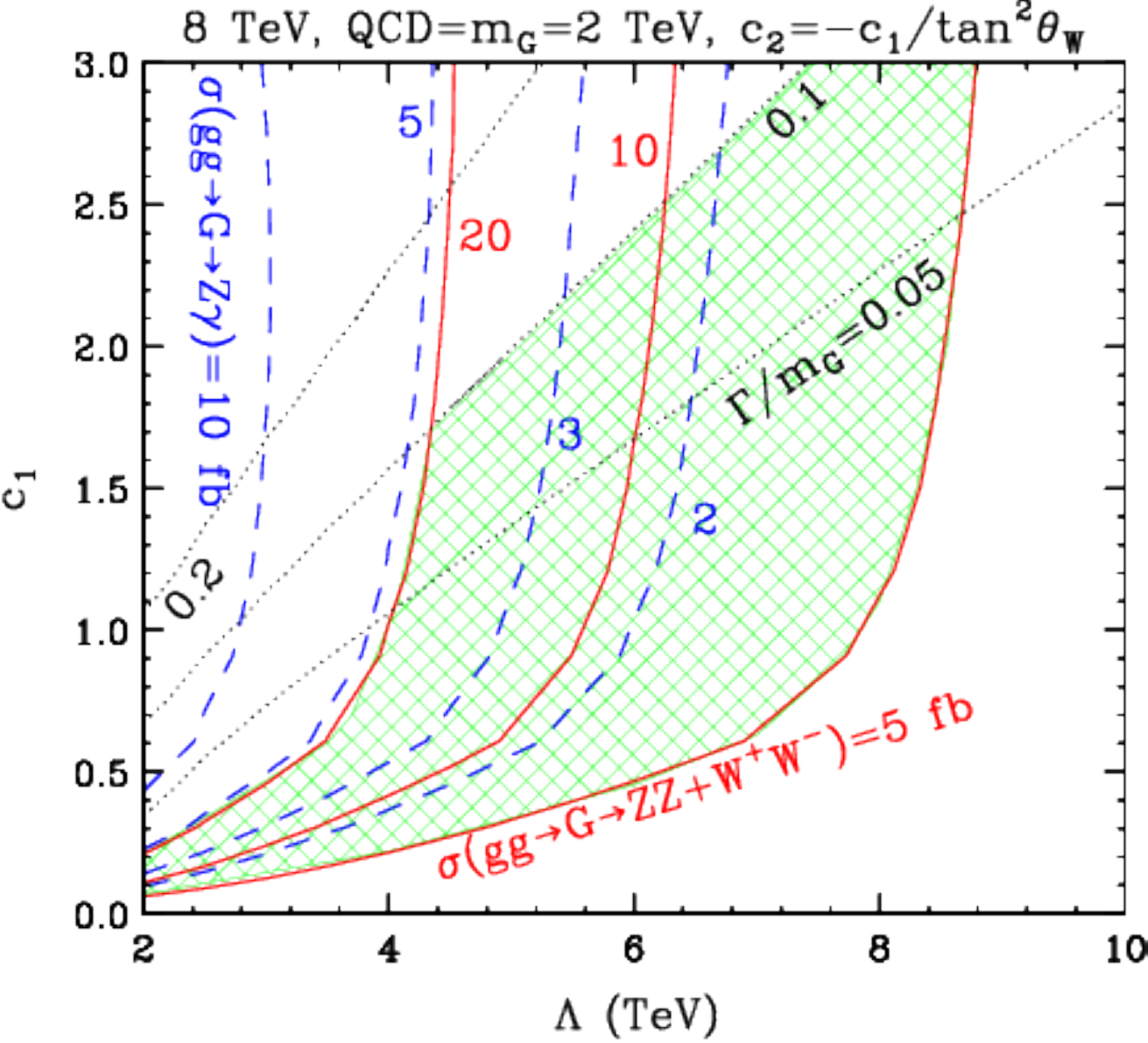}
 }
\caption{Similar to Fig. \ref{fig:vector} but for the CP-even tensor. } 
\label{fig:tensor}
\end{center}
\vspace*{-0.5cm}
\end{figure}

The partial decay widths of the spin-2 resonance with CP-even interactions into $\gamma\gamma$, $Z\gamma$, $ZZ$, $W^+W^-$, and $gg$~\cite{GMDM} are 
%\be
%\Gamma_{\cal G}= \Gamma_{\cal G}(\gamma\gamma)+\Gamma_{\cal G}(Z\gamma)+\Gamma_{\cal G}(ZZ)+\Gamma_{\cal G}(W^+ W^-)+\Gamma_{\cal G}(gg) \, ,
%\ee
%with 
%
\bea
\begin{cases}
\Gamma_{\cal G}(\gamma\gamma)= \frac{ |c_{\gamma\gamma}|^2 m^3_G}{80\pi\Lambda^2}\, , 
&c_{\gamma\gamma} = c_1 \cos^2\theta_W + c_2 \sin^2\theta_W\\
\Gamma_{\cal G}(Z Z)= \frac{|c_{ZZ}|^2 m^3_G}{80 \pi \Lambda^2} 
			\sqrt{1 - 4x^G_Z} \left (  1- 3x^G_Z  + 6(x^G_Z)^2 \right)     \, ,
			&c_{ZZ} = c_2 \cos^2\theta_W+c_1 \sin^2\theta_W\\
\Gamma_{\cal G}(Z \gamma)= \frac{|c_{\gamma Z}|^2 m^3_G}{160 \pi \Lambda^2}  
			\left ( 1 - {x^G_Z} \right )^3  \left (  1+ \frac{1}{2}x^G_Z +\frac{1}{6} (x^G_Z)^2 \right)     \, ,
			&c_{Z\gamma} = (c_2-c_1)\sin(2\theta_W)\\
\Gamma_{\cal G}(W^+ W^- )= \frac{|c_{WW}|^2 m^3_G}{160 \pi \Lambda^2} 
			\sqrt{1 - 4 x^G_W} \left (1- 3x^G_W  + 6(x^G_W)^2 \right)    \, , 
			&c_{WW}=2 c_2\\
\Gamma_{\cal G}(gg)=  \frac{|c_{gg}|^2 m^3_G}{10\pi\Lambda^2} \,  ,
&c_{gg}=c_3 \, ,
\end{cases}% \nonumber \\
%\vspace{-5cm} 
\eea
where again $x^G_i$ is defined in Eq.~(\ref{eq:ratio}).
One may notice that for $c_1 =  c_2$, the decay mode, ${\cal G}\rightarrow Z\gamma$, vanishes.
We note that the branching fractions of the spin-2 resonance are of the similar form as the ones of scalar resonances discussed in Section~\ref{sec:scalar} because the spin-2 resonance decays through gauge invariant operators composed of field strength tensors. One can also suppress the diphoton rate by choosing $c_2 = -c_1/\tan^2\theta_W$, 
which forces all relevant branching fractions to be the same as those in the scalar case.
Our parameter scan results are summarized in Fig.~\ref{fig:tensor}. 
We also find that production cross section, $gg \to {\cal G} \to gg$, in the demonstrated parameter space was small and therefore, there is no constraint from the LHC dijet resonance search.

%%%%%%%%%%%% 
%%%%%%%%%%%%% 

\red{On the other hand, there is no counterpart of the energy-momentum tensor for CP-odd interactions, but the Lorentz invariance and the gauge invariance dictate the detailed form of the interactions.  
Following the similar step as in the CP-even vector case, the CP-odd interactions of the spin-2 resonance to the SM gauge bosons\footnote{The $ZZ$ coupling to the CP-odd tensor field was considered in Ref.~\cite{gao,Avery:2012um} without gauge invariance imposed. } are given by
\be
{\cal L}^{\cal G}_{\rm CP-odd}=\frac{1}{\Lambda} {\cal G}_{\mu\nu} {\tilde T}^{\mu\nu} \, ,
\ee
where 
\bea
{\tilde T}_{\mu\nu}&=& a_1 \,\epsilon_{\mu\lambda\rho\sigma}  \partial^\lambda Z_\nu  {F}^{\rho\sigma}
+{\hat a}_1\,  \epsilon_{\mu\lambda\rho\sigma}\partial^\lambda Z_\nu   (\partial^\rho Z^\sigma - \partial^\sigma Z^\rho) \nonumber \\
&&+\Big(a_2\,\epsilon_{\mu\lambda\rho\sigma}\partial^\lambda W^-_\nu ( \partial^\rho W^{\sigma +} -\partial^\sigma W^{\rho +})+i{\hat a}_2 \partial^\lambda W^-_\nu( \partial_\mu W^{+}_\lambda -\partial_\lambda W^{+}_\mu) +{\rm c.c.}\Big)  \nonumber \\
&&+\frac{a_3}{\Lambda}\, \partial^\lambda\partial_\nu h\,{F}_{\mu\lambda} +\frac{{\hat a}_3}{\Lambda} \, \partial^\lambda\partial_\nu h\,(\partial_\mu Z_\lambda-\partial_\lambda Z_\mu).
\eea
Here $a_2$ and  ${\hat a}_2 $ are related to other parameters through gauge invariance as 
\be
a_2 = \frac{m_W}{m_Z}\Big(a_1 \sin\theta_W +{\hat a}_1 \cos\theta_W \Big), \quad
{\hat a}_2 = -\frac{m_W}{\Lambda}\Big(a_3 \sin\theta_W+{\hat a}_3 \cos\theta_W \Big).
\ee
The operators in ${\tilde T}_{\mu\nu}$ are induced from higher dimensional gauge-invariant operators such as $[D^\lambda D_\nu H]^\dagger {\tilde F}^Y_{\mu\lambda} H$,\; $[D^\lambda D_\nu H]^\dagger {F}^Y_{\mu\lambda} H$,\; $[D^\lambda D_\nu H]^\dagger {\tilde F}^W_{\mu\lambda} H$,  and $[D^\lambda D_\nu H]^\dagger {F}^W_{\mu\lambda} H$ after electroweak symmetry breaking.\footnote{We note that one of higher dimensional operators among $[D_\nu D^\lambda H]^\dagger {\tilde F}^Y_{\mu\lambda} H$ and $[D^\lambda D_\nu H]^\dagger {\tilde F}^Y_{\mu\lambda} H$ is redundant because $[D^\lambda D_\nu H-D_\nu D^\lambda H]^\dagger {\tilde F}^Y_{\mu\lambda} H\sim |H|^2 F^{Y\lambda}\,_\nu {\tilde F}^Y_{\mu\lambda} $, which contributes to the gauge invariant operators of $F^{Y\lambda}\,_\nu {\tilde F}^Y_{\mu\lambda} $ that becomes a vanishing gauge interaction after electroweak symmetry breaking (EWSB).} Therefore, the resulting effective CP-odd interactions of the spin-2 resonance are of strong similarity to those of the spin-1 resonances as discussed in Section~\ref{sec:vector}.
Hence, the spin-2 resonance can decay into a pair of electroweak gauge bosons or Higgs bosons via symmetry breaking terms in ${\tilde T}_{\mu\nu}$. We note that as in the CP-odd vector case, the unitarity cutoff of $\Lambda\sim 10\,{\rm TeV}$ implies that $a_{1,2},\; {\hat a}_{1,2} \lesssim {\cal O}(10^{-2})$ and  $a_3,\; {\hat a}_3\lesssim {\cal O}(1)$.}

\red{We note, however, that diquark CP-odd operator, ${\bar q}\gamma^5(\gamma^\mu\partial^\nu+\gamma^\nu\partial^\mu) q +{{\rm h.c.}}$, is a total derivative, while a nontrivial diquark operator, $i{\bar q}\gamma^5(\gamma^\mu\partial^\nu+\gamma^\nu\partial^\mu) q+{\rm h.c.} $ is CP-even \cite{gersdorff}. 
Since the CP-odd diquark operator can be written as $(\partial^\nu {\cal G}_{\mu\nu}){\bar q}\gamma^5 \gamma^\mu q$ by integration by parts, the diquark production of the on-shell CP-odd spin-2 resonance is suppressed due to $\partial^\nu {\cal G}_{\mu\nu}=0$.
Instead, the CP-odd spin-2 resonance can be produced by vector boson fusion. In this case, there are two forward jets accompanying the resonance, so we cannot explain the ATLAS diboson excess by the CP-odd spin-2 resonance. For this reason, we do not consider it any longer.  
We also remark that the operators composed of field strength tensors only, for example, ${\cal  G}_{\mu\nu}{\rm Tr}({\tilde F}_{\mu\lambda}F^{\lambda}\,_\nu)$ with $F_{\mu\nu}$ being $F^Y_{\mu\nu}$, $F^W_{\mu\nu}$, or $G_{\mu\nu}$, vanish because ${\cal  G}_{\mu\nu}{\rm Tr}({\tilde F}_{\mu\lambda}F^{\lambda}\,_\nu)=-\frac{1}{4}{\cal G}_\mu^\mu\,{\rm Tr}(F_{\alpha\beta}{\tilde F}^{\alpha\beta})=0$ due to the traceless condition, i.e., ${\cal G}^\mu_\mu=0$. Therefore, those gauge invariant operators do not contribute to the process with on-shell CP-odd tensor so that the gluon fusion production of the CP-odd spin-2 resonance is suppressed.
For a future reference on the phenomenological study of the CP-odd spin-2 resonance, we summarize the partial decay rates with CP-odd interactions into $Z\gamma$, $ZZ$, $W^+W-$, $h\gamma$, $hZ$ and $q\bar{q}$ as listed below:
%
%\be
%\Gamma_{\cal G}=\Gamma_{\cal G}(Zh)+\Gamma_{\cal G}(\gamma h)+ \Gamma_{\cal G}({\bar q}q)+\Gamma_{\cal G}(ZZ)+\Gamma_{\cal G}(W^+ W^-)+\Gamma_{\cal G}(Z\gamma)
%\ee
%with
\begin{eqnarray}
\left\{
\begin{array}{l}
\Gamma_{\cal G}( Z \gamma)= \frac{ a_1^2 m_G^3}{960  \pi \Lambda^2} \left (1-x_Z^G \right )^3 \left ( 34 + 3 x_Z^G + 3 (x_Z^G)^{-1}\right ), \\
\Gamma_{\cal G}( Z Z)= \frac{ \hat a_1^2 m_G^5}{ 960 \pi m_Z^2  \Lambda^2} \sqrt{ 1- 4x_Z^G}  \left ( 3 -4 x_Z^G-32 (x_Z^G)^2 \right )    \, , \\
\Gamma_{\cal G}( W^+W^-)=  \frac{ m_G^5}{1920 \pi m_W^2 \Lambda^2} \left ( 1- 4x_W^G \right )^{3/2} \left [ 3 \hat a_2^2 \left ( 1- 4x_W^G \right ) + 4 a_2^2 \left ( 3 +  8x_W^G \right )\right ]    \, ,\\
\Gamma_{\cal G}( h\gamma )=  \frac{a_3^2 \, m_G^5}{1280 \pi \Lambda^4} \left ( 1 - x_h^G \right )^5    \, , \\
\Gamma_{\cal G}( hZ)=   \frac{ \hat a_3^2 \, m_G^5}{3840 \pi \Lambda^4} 
 \left [  \Big (1 -  x_h^G \Big )^2 - 2 \Big (1+ x_Z^G \Big )  x_Z^G + (x_Z^G)^2 \right ]^{5/2}        \\
\hspace{1.5cm}\times \left [      3 \Big ( 1 -  x_Z^G\Big )^2 +  2 \Big ( -2 + 5  x_h^G +  (x_h^G)^2  \Big )  x_Z^G - \Big ( 1+4  x_h^G \Big )  (x_Z^G)^2 + 2  (x_Z^G)^3
\right ]
 \\
\hspace{1.5cm}\approx  \frac{ \hat a_3^2 \, m_G^5}{1280 \pi \Lambda^4}  \, 
%\Gamma_{\cal G}(q \bar{q})= \frac{3 a_q^2 m_G^3}{40 \pi \Lambda^2}  \left ( 1- x_q^G  \right )^{5/2}    \, , \\
\end{array}\right.
\end{eqnarray}
where again $x_i^G$ is defined in Eq.~(\ref{eq:ratio}).  }
%

%%%%%%%%%%%%%%%%%%%%%%%%%%%%%%%%%%%%%%%%%%%%%%%%%%%%
\section{Kinematic Correlations in the Diboson Final State} \label{sec:kinematics}
%%%%%%%%%%%%%%%%%%%%%%%%%%%%%%%%%%%%%%%%%%%%%%%%%%%%

In this section, we discuss ways of discriminating potential scenarios to give rise to diboson resonances. Since we have observed that various bosonic particles with different spins and CP states can accommodate the excesses reported by the ATLAS collaboration with a suitable choice of parameters, it is of paramount importance to pin down the underlying physics once those excesses are confirmed experimentally. Of potentially useful variables, we employ several angular correlations between the decay products of the resonance of interest. We first suppose that a resonance $R$ decays into two vector bosons $V_1$ and $V_2$ which subsequently decay into two visible particles $u_i$ and $v_i$ $(i=1,2)$:
\bea
pp\rightarrow R \rightarrow V_1(\rightarrow u_1+v_1)+V_2(\rightarrow u_2+v_2),
\eea
and denote $\vec{P}_i$ as the three momentum of $V_i$ and $\vec{u}_i(\vec{v}_i)$ as those of $u_i (v_i)$. 

With these notations, we enumerate the angular variables to be used here as follows:
\bea
\Phi&=&\frac{\vec{P}_1\cdot(\hat{n}_1\times \hat{n}_2)}{|\vec{P}_1\cdot(\hat{n}_1\times \hat{n}_2)|}\cos^{-1}(\hat{n}_1\cdot \hat{n}_2) \hbox{ with } \hat{n}_i=\frac{\vec{u}_i \times \vec{v}_i}{|\vec{u}_i \times \vec{v}_i|}, \\
\Phi_1&=&\frac{\vec{P}_1\cdot(\hat{n}_1\times \hat{n}_{sc})}{|\vec{P}_1\cdot(\hat{n}_1\times \hat{n}_{sc})|}\cos^{-1}(\hat{n}_1\cdot \hat{n}_{sc}) \hbox{ with } \hat{n}_{sc}=\frac{\hat{z} \times \vec{P}_1}{|\hat{z} \times \vec{P}_1|}, \\
\cos\theta^* &=& \frac{\vec{P}_1\cdot \hat{z}}{|\vec{P}_1|}, \\
\cos\theta_1 &=& -\frac{\vec{P}_2\cdot \vec{u}_1}{|\vec{P}_2||\vec{u}_1|}.
\eea
For the first three variables, all the momenta are measured in the rest frame of resonance $R$, while for the last one, all the momenta are measured in the rest frame of vector boson $V_1$. These variables have been used in the context of resonance discrimination in the literature, and they show distinctive structures depending on quantum numbers of each resonance (see, for example, Ref.~\cite{gao}). 

\begin{table}[t]
\centering
\begin{tabular}{c| c |c}
\hline \hline
Scenario & Parameter choice & $R$ production \\
\hline
$0^+$ & $s_1=0.4,\;s_2=-s_1/\tan^2\theta_W,\;s_3=1,\; \Lambda=10$ TeV & $gg\rightarrow R$\\
$0^-$ & $a_1=0.6,\;a_2=-a_1/\tan^2\theta_W,\;a_3=1,\;\Lambda=20$ TeV & $gg\rightarrow R$\\
$1^+$ & $\hat{\kappa}_1=0.008,\;g_X=0.02,\;c_L=c_R=1,\;\Lambda=10$ TeV & $q\bar{q}\rightarrow R$\\
$1^-$ & $\hat{\alpha}_1=0.01,\;g_X=0.04,\;-c_L=c_R=1,\; \Lambda=10$ TeV & $q\bar{q}\rightarrow R$\\
$2^+$ & $c_1=0.5,\;c_2=-c_1/\tan^2\theta_W,\;c_3=1,\;\Lambda=5$ TeV & $gg\rightarrow R$\\
% \\$2^-$ & $\hat{a}_1=0.1,\; a_q=0.25,\;\Lambda=10$ TeV & $q\bar{q}\rightarrow R$\\
\hline \hline
\end{tabular}
\caption{\label{tab:paramChoice} List of scenario choices for a resonance $R$ having a spin and CP-state denoted as $J^{CP}$. }
\end{table}

\begin{figure}[t]
\begin{center}
\centerline{
\includegraphics[width=7.7cm]{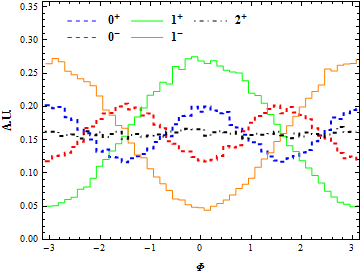}\hspace{0.5cm}
\includegraphics[width=7.7cm]{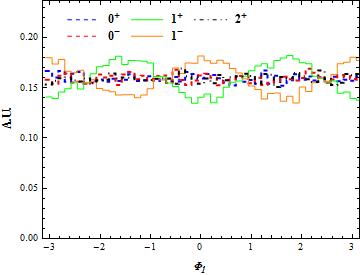} \vspace{0.5cm} } 
\centerline{
\includegraphics[width=7.7cm]{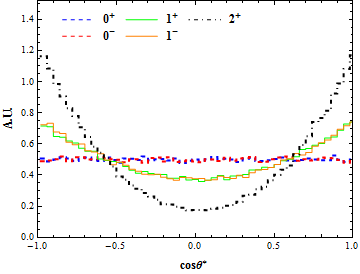} \hspace{0.5cm}
\includegraphics[width=7.7cm]{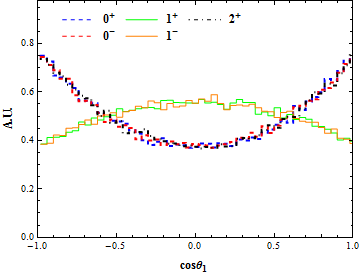}}
\caption{\label{fig:variables} Unit-normalized distributions in $\Phi$ (upper left panel), $\Phi_1$ (upper right panel), $\cos\theta^*$ (lower left panel), and $\cos\theta_1$ (lower right panel) for the resonance decay into two $W$ gauge bosons. The spin and CP state of the resonance of interest is represented by $J^{CP}$.}
\end{center}
\vspace{-0.5cm}
\end{figure}

We here study the above-listed observables in the analysis of $R\rightarrow W^+W^-$, and show the distributions in Fig. \ref{fig:variables}. The distributions are plotted with parton-level events with a 10\% of Gaussian smearing onto energy of each final state particle for more realistic Monte Carlo simulation. Again, events were generated by \texttt{MadGraph\_aMC@NLO}~\cite{Alwall:2014hca} together with the default set of parton distributions \texttt{NNPDF23}~\cite{Ball:2012cx} at the center of mass energy of 13 TeV. Table~\ref{tab:paramChoice} summarizes our parameter choices for each scenario. All the parameters {\it not} listed in the table are simply taken to be zero. 
Note that this choice of parameters is made only for the purpose of illustration of different kinematic distributions for each scenario. We find that the shape is not strongly dependent on parameters.
The mass and the total decay width are fixed to be 2 TeV and 0.1 TeV, correspondingly. We remark that the spin-0 and CP-even spin-2 resonances are produced via gluon fusion while the spin-1 resonances are produced via quark annihilation. The observables of $\Phi$, $\Phi_1$, $\cos\theta^*$, and $\cos\theta_1$ are exhibited in the upper left panel, the upper right panel, the lower left panel, and the lower right panel, respectively. Different spin and CP states are symbolized by $J^{CP}$, and they are histogramed as follows: CP-even scalar by the blue dashed, CP-odd scalar by the red dashed, CP-even vector by the green solid, CP-odd vector by the orange solid, and CP-even tensor by the black dot-dashed. 
In particular, the theory prediction for $\cos\theta^*$ distributions is readily derived as follows: 
\begin{equation}
\frac{d \sigma}{d \cos\theta^*} \sim \left \{
\begin{array}{ll}
1 \, ,                                                     & \hspace{1cm}  \textrm{for }  g g \to 0^+ \, , 0^-   \to W^+ W^-\\
1 + \cos^2\theta^* \, ,                              &  \hspace{1cm}   \textrm{for } q \bar{q} \to  1^+ \, ,1^-   \to W^+ W^-\\  
1 + 6 \cos^2\theta^* + \cos^4\theta^* \, ,    &  \hspace{1cm}  \textrm{for } g g \to ~~ \, 2^+ \, ~~\to W^+ W^-
%1 - 3 \cos^2\theta^* + 4 \cos^4\theta^* \, ,    &  \hspace{1cm}  \textrm{for } q {\bar q} \to ~~\,  2^- \, ~~\to W^+ W^-\\
\end{array}
\right. \, .
\label{eq:costheta}
\end{equation}
which can be directly compared with experimental data. 

First of all, we observe that the angular distributions with the Gaussian smearing are very similar to those without any smearing, from which we expect that the angular distributions are insensitive to detector effects such as jet energy resolution. Moving onto Fig.~\ref{fig:variables}, we clearly see that these observables are useful enough to distinguish potential scenarios associated with diboson resonances. For example, the CP-even vector resonance (green solid histograms) shows distinctive behaviors in all four observables. Furthermore, the unique features according to different spin and CP states in those variables can be used for cross-checks. 
Note that one single distribution can not discriminate different scenarios, and thus it is important to consider all possible kinematic correlations.
Finally, we remark that similar analyses can be straightforwardly applicable to other diboson resonances such as $R \rightarrow h \gamma$, $R\rightarrow hZ$ and $R\rightarrow Z\gamma$  so that more information can be extracted to confirm the underlying physics governing the observed phenomena.

%%%%%%%%%%%%%%%%%%%%%%%%%%%%%%%%%%%%%%%%%%%%%%%%%%%%%%%%%%%%
\section{Summary} \label{sec:discussion}
%%%%%%%%%%%%%%%%%%%%%%%%%%%%%%%%%%%%%%%%%%%%%%%%%%%%%%%%%%%%

Recently, the ATLAS collaboration has reported some excesses in searches for diboson resonances using jet-substructure techniques. 
The excesses show up in the invariant mass of $W^+W^-$, $W^\pm Z$ and $ZZ$ at around 2 TeV.
It has been discussed in literature that about 20\% of the events in at least one signal region belong to all three categories, 
which indicates that these ``resonances'' may be explained by one single particle rather than two.

In this paper, we have explored a possible new physics interpretation of the ATLAS diboson excess in an effective field theory approach, which covers a rather large class of models in a reasonably model independent manner. 
We considered the effective operators for scalar ($s=0$), vector ($s=1$), and tensor ($s=2$) resonances with different CP properties. It is shown that each scenario may explain the ATLAS diboson excess without contradicting other constraints, except the CP-odd spin-2 resonance whose diquark or gluon fusion production is suppressed. The CP-odd vector case might have some tension with the unitary bound. 
Symmetries of each scenario predict signals in other final states such as $Z\gamma$ and $\gamma\gamma$ in the cases of scalar and CP-even tensor resonances; $Z\gamma$ and $hZ$, $h \gamma$ at a smaller rate in the cases of vector resonances. 
Especially, the dijet, $t\bar{t}$, $Z\gamma$, $hZ$, and $h \gamma$ resonance searches at the LHC run II may confirm or constrain these scenarios. 

With limited statistics, all these scenarios provide a relatively good fit to the data.
However, a further accumulation of data might reveal the real identity of the resonance.
We showed a few examples of kinematic distributions, which are sensitive to the CP property and spin of the resonance. We strongly encourage experimental collaborations to look at these kinematic correlations.

%%%%%%%%%%%%%%%%%%%%%%%%%%%%%%%%%%%%%%%%%%%%%%%%%%%%%%%%%%%%
%\newpage
%\begin{acknowledgements}
\section*{Acknowledgments} %JHEP style

We thank the Center for Theoretical Underground Physics and Related Areas (CETUP* 2015) for hospitality and partial support during the completion of this work. 
Especially we are grateful to Barbara Szczerbinska for all the arrangements and encouragment.
We also thank Kingman Cheung, Pyungwon Ko, Jong-Chul Park, Veronica Sanz and  Yeo Woong Yoon for helpful discussion. 
This work is partially supported by the Basic Science Research Program through the National Research Foundation of Korea funded by the Ministry of Education, Science and Technology (2013R1A1A2064120 and 2013R1A1A2007919).  
D.~K. is supported by the LHC Theory Initiative postdoctoral fellowship (NSF Grant No. PHY-0969510), and K.~K. is supported by the U.S. DOE under Grant No. DE-FG02-12ER41809.

%\end{acknowledgements}
%%%%%%%%%%%%%%%%%%%%%%%%%%%%%%%%%%%%%%%%%%%%%%%%%%%%%%%%%%%%

%%%%%%%%%%%%%%%%%%%%%%%%%%%%%%%%%%%%%%%%%%
%\newpage
%\section*{Appendix}

%\newpage
\appendix
\section{Decay widths}
In this appendix, we summarize the useful formulas for the decay rates for scalar and tensor resonances.

\subsection{CP-even scalar}
For interaction Lagrangian ${\cal L}= -c_{V_1V_2}\frac{\phi}{\Lambda}F_{V_1}^{\mu\nu}F_{V_2 \mu\nu}$, the decay width of $\phi$ to $V_1V_2$ is given as
\begin{eqnarray}
&&\Gamma(\phi\to V_1V_2) = \frac{s_V|c_{V_1V_2}|^2}{8\pi}\cdot \left(\frac{m_\phi^3}{\Lambda^2}\right)\cdot {\cal F}(\frac{m_1}{m_\phi},\frac{m_2}{m_\phi}),
\end{eqnarray}
where ${\cal F}(x_1,x_2)$ is defined as 
\begin{eqnarray}
&&{\cal F}(x_1,x_2)=\left(1-(x_1+x_2)^2\right)^{1/2}\left(1-(x_1-x_2)^2\right)^{1/2}\left(1+x_1^4+x_2^4-2(x_1^2+x_2^2)+4x_1^2 x_2^2\right) \, .\nonumber \\
\end{eqnarray}
$s_V$ is symmetric factor, which is 1 for $V_1\neq V_2$ and $2$ for $V_1=V_2$, respectively. 

\subsection{CP-odd scalar} 

For interaction Lagrangian ${\cal L}= -c_{V_1V_2}\frac{A}{\Lambda}F_{V_1}^{\mu\nu}\tilde{F}_{V_2 \mu\nu}$, the decay width of $A$ to $V_1V_2$ is given as
\begin{eqnarray}
&&\Gamma(A\to V_1V_2) = \frac{s_V|c_{V_1V_2}|^2}{2\pi}\cdot \left(\frac{m_A^3}{\Lambda^2}\right)\cdot {\cal G}(\frac{m_1}{m_A},\frac{m_2}{m_A}),\\
&&{\cal G}(x_1,x_2)=\left(1-(x_1+x_2)^2\right)^{3/2}\left(1-(x_1-x_2)^2\right)^{3/2}
\end{eqnarray}
where $s_V$ is symmetric factor, which is 1 for $V_1\neq V_2$ and $2$ for $V_1=V_2$, respectively.

\subsection{CP-even tensor}

For spin-2 tensor with mass $m_G$, the interaction Lagrangian is ${\cal L} = -c_{V_1 V_2} \frac{{\cal G}_{\mu\nu}}{\Lambda}{F_{V_1}}^\mu_{~\lambda} {F_{V_2}}^{\lambda\nu}$ and the decay width of ${\cal G}_{\mu\nu} \to V_1 V_2$ is given as
\begin{eqnarray}
\Gamma({\cal G}_{\mu\nu}\to V_1 V_2) = \frac{s_V|c_{V_1V_2}|^2 m_G^3}{160\pi \Lambda^2}{\cal H}(\frac{m_1}{m_h},\frac{m_2}{m_h}),
\end{eqnarray}
where the convenient dimensionless function, ${\cal H}(x,y)$, for some interesting cases are 
\begin{eqnarray}
{\cal H}(x,x) = \sqrt{1-4x^2}(1-3x^2+6x^4),\\
{\cal H}(x,0)= (1-x^2)^3(1+\tfrac{1}{2}x^2 +\tfrac{1}{6}x^4)
\end{eqnarray}
and the symmetric factor  $s_V$ is 1 for $V_1\neq V_2$ and $2$ for $V_1=V_2$, respectively.

%%%%%%%%%%%%%%%%%%%%%%%%%%%%%%%%%%%%%%%%%%%%%%%%%%%%%%%%%%%%

\end{document}